\documentclass[journal,final,onecolumn,12pt,twoside]{IEEEtran}
\usepackage{times,amssymb,amsbsy,amsmath,amsthm,epsfig,nicefrac,euscript,mathrsfs,mathrsfs,color}
\usepackage{epstopdf}
\usepackage{amsmath,amsthm}
\usepackage{enumerate}
\usepackage{eufrak}
\usepackage{mathcomp}
\usepackage{supertabular}
\usepackage{longtable}
\usepackage{stmaryrd}
\usepackage{url}
\usepackage{color}
\usepackage{rotating}
\usepackage{float}
\usepackage{array}
\usepackage{mathtools}
\usepackage{multicol}
\usepackage{algorithm}
\usepackage{algorithmic}
\usepackage{xcolor}
\usepackage[all]{xy}
\usepackage{wrapfig}
\usepackage{graphicx}
\usepackage[font=small]{caption}
\usepackage{subcaption}
\usepackage{xspace,exscale,relsize}
\usepackage{fancybox,shadow}
\usepackage{color}
\usepackage{amsfonts}
\usepackage{url}
\usepackage[all]{xy}
\usepackage[noadjust]{cite}
\usepackage{rotating}
\usepackage{ifpdf}
\usepackage{psfrag}
\usepackage{eulervm} % a better implementation of the euler package (not in gwTeX)

\newcommand{\remove}[1]{}
\newcommand{\poly}{\textnormal{poly}}

\newcommand{\nc}{\newcommand}
\newcommand{\vA}{{\bf A}}

\newcommand{\vB}{{\bf B}}

\newcommand{\vH}{{\bf H}}

\newcommand{\vb}{{\bf b}}

\newcommand{\vu}{{\bf u}}
\newcommand{\vv}{{\bf v}}

\newcommand{\vhx}{{\widehat{\bf x}}}

\newcommand{\vzero}{{\bf 0}}
\newcommand{\vone}{{\bf 1}}
\newcommand{\vbeta}{{\boldsymbol \beta}}

\newcommand{\A}{\mathcal A}
\newcommand{\B}{\mathcal B}
\newcommand{\C}{\mathcal C}

\newcommand{\E}{\mathcal E}
\newcommand{\F}{\mathcal F}

\newcommand{\Q}{\mathcal Q}
\newcommand{\Qb}{\bar{\mathcal Q}}

\newcommand{\pQ}{{\bf p}\mathcal Q}
\newcommand{\pQb}{{\bf p}\bar{\mathcal Q}}

\newcommand{\ZZ}{\mathbb Z}

\newcommand{\bbracket}[1]{\left\llbracket{#1}\right\rrbracket}

\def\beq{\begin{equation}}
\def\eeq{\end{equation}}
\nc\half{\nicefrac12}

\nc\bfa{{\boldsymbol a}}\nc\bfA{{\mathbf A}}\nc\cA{{\mathcal A}}
\nc\bfb{{\boldsymbol b}}\nc\bfB{{\mathbf B}}\nc\cB{{\mathcal B}}
\nc\bfc{{\boldsymbol c}}\nc\bfC{{\mathbf C}}\nc\cC{{\mathcal C}}
\nc\bfd{{\boldsymbol d}}\nc\bfD{{\mathbf D}}\nc\cD{{\mathcal D}}\nc\sD{{\mathscr D}}
\nc\bfe{{\boldsymbol e}}\nc\bfE{{\mathbf E}}\nc\cE{{\EuScript E}}
\nc\bff{{\boldsymbol f}}\nc\bfF{{\mathbf F}}\nc\cF{{\mathcal F}}
\nc\bfg{{\boldsymbol g}}\nc\bfG{{\mathbf G}}\nc\cG{{\mathcal G}}
\nc\bfh{{\boldsymbol h}}\nc\bfH{{\mathbf H}}\nc\cH{{\mathcal H}}
\nc\bfi{{\boldsymbol i}}\nc\bfI{{\mathbf I}}\nc\cI{{\mathcal I}}
\nc\bfj{{\boldsymbol j}}\nc\bfJ{{\mathbf J}}\nc\cJ{{\mathcal J}}
\nc\bfk{{\boldsymbol k}}\nc\bfK{{\mathbf K}}\nc\cK{{\mathcal K}}
\nc\bfl{{\boldsymbol l}}\nc\bfL{{\mathbf L}}\nc\cL{{\mathcal L}}\nc\sL{{\mathscr L}}
\nc\bfm{{\boldsymbol m}}\nc\bfM{{\mathbf M}}\nc\cM{{\mathcal M}}
\nc\bfn{{\boldsymbol n}}\nc\bfN{{\mathbf N}}\nc\cN{{\mathcal N}}
\nc\bfo{{\boldsymbol o}}\nc\bfO{{\mathbf O}}\nc\cO{{\mathcal O}}
\nc\bfp{{\boldsymbol p}}\nc\bfP{{\mathbf P}}\nc\cP{{\mathcal P}}
\nc\bfq{{\boldsymbol q}}\nc\bfQ{{\mathbf Q}}\nc\cQ{{\mathcal Q}}\nc\sQ{{\mathscr Q}}
\nc\bfr{{\boldsymbol r}}\nc\bfR{{\mathbf R}}\nc\cR{{\mathcal R}}
\nc\bfs{{\boldsymbol s}}\nc\bfS{{\mathbf S}}\nc\cS{{\mathcal S}}
\nc\bft{{\boldsymbol t}}\nc\bfT{{\mathbf T}}\nc\cT{{\mathcal T}}\nc\sT{{\mathscr T}}
\nc\bfu{{\boldsymbol u}}\nc\bfU{{\mathbf U}}\nc\cU{{\mathcal U}}
\nc\bfv{{\boldsymbol v}}\nc\bfV{{\mathbf V}}\nc\cV{{\mathcal V}}
\nc\bfw{{\boldsymbol w}}\nc\bfW{{\mathbf W}}\nc\cW{{\mathcal W}}\nc\sW{{\mathscr W}}
\nc\bfx{{\boldsymbol x}}\nc\bfX{{\mathbf Z}}\nc\cX{{\EuScript X}}
\nc\bfy{{\boldsymbol y}}\nc\bfY{{\mathbf Y}}\nc\cY{{\EuScript Y}}\nc\sY{{\mathscr Y}}
\nc\bfz{{\boldsymbol z}}\nc\bfZ{{\mathbf Z}}\nc\cZ{{\mathcal Z}}\nc\sZ{{\mathscr Z}}

\def\path{{\text{path}}}

\newcommand{\define}{\stackrel{\mbox{\tiny $\triangle$}}{=}}
\newcommand{\X}{{\mathcal X}}
\newcommand{\fq}{\mathbb{F}_q}

\def\h_q{\qopname\relax{no}{h_q}}
\def\h{\qopname\relax{no}{h}}

\newcounter{arb}

\newcommand{\beas}{\begin{eqnarray*}} 
\newcommand{\eeas}{\end{eqnarray*}} 

\begin{document}

\title{DNA-Based Data Storage Systems: A Review of Implementations and Code Constructions}

\author{Olgica~Milenkovic~and~Chao~Pan
\thanks{O. Milenkovic and C. Pan are with the Department
of Electrical and Computer Engineering, University of Illinois, Urbana-Champaign, Urbana, IL, USA. e-mail: milenkov@illinois.edu, chaopan2@illinois.edu.}}% <-this % stops a space

\markboth{DNA-Based Data Storage Systems: A Review of Implementations and Code Constructions}%
{DNA-Based Data Storage Systems: A Review of Implementations and Code Constructions}

\maketitle

\vspace{-0.3in}
{\small \begin{abstract}
In this review paper, we delve into the nascent field of molecular data storage, focusing on system implementations and code constructions. We start by providing an overview of basic concepts in synthetic and computational biology. Afterwards, we proceed with a review of the diverse approaches followed to implement such systems. In the process, we identify new problems in communication and coding theory, and discuss some relevant results pertaining to DNA sequence profiles, coded trace reconstruction, coding for DNA punchcard systems and coding for unique reconstruction.
\end{abstract}}
% \fi
\IEEEpeerreviewmaketitle
\vspace{-0.1in}
\section{Introduction and Motivation}
%%%%%%%%%%%%%%%%%%%%%%%%%%%%%%%%%%%%%%%%%%%%%%%%%
Despite numerous advancements in traditional data recording techniques, the emergence of Big Data platforms and the growing concern for energy conservation have presented challenges for the storage community to develop new nonvolatile, durable storage media that can handle ultrahigh volumes of data. 

The potential use of macromolecules for data storage was recognized as far back as the 1960s when Richard Feynman outlined his nanotechnology vision in his talk ``There is plenty of room at the bottom''~\cite{feynman1960there}. Among the various macromolecules that could potentially serve as storage media, DNA molecules are particularly promising due to their unique properties. DNA has been successfully used as a building block for small-scale self-assembly-based computers~\cite{seeman2007overview}. 
It is also capable of preserving its contents under harsh weather conditions for thousands of years, as demonstrated by the recovery of DNA from $30,000$ years old Neanderthal and $700,000$ years old horse bones~\cite{saey2013story}. In addition, DNA allows for extremely high storage capacities, with a single human cell containing DNA strands that encode $6.4$ gigabits of information within a mass of only approximately $3$ picograms. The technologies for DNA amplification and synthesis have also reached unprecedented levels of efficiency and accuracy~\cite{shendure2012expanding}, while DNA sequencing has been a standard procedure for nearly two decades.

Building on the progress of DNA synthesis and sequencing technologies, two laboratories described the first architectures for archival DNA-based storage in 2012~\cite{church2012next,goldman2013towards}. The first architecture achieved a density of $700$ TB/gram, while the second approach improved the density to $2$ PB/gram. The superior density of the second approach may be attributed in part to the use of basic coding schemes such as Huffman coding, runlength coding, single parity-check coding, and repetition coding. Subsequent works~\cite{grass2015robust} extended the coding approach of the second architecture to account for missing DNA fragments via Reed-Solomon codes~\cite{reed1960polynomial}. 

Further milestones in DNA-based data storage were reached through several innovations. The first innovation was the introduction of random-access and rewriting platforms enabled by controlled polymerase chain reaction (PCR) and/or overlap-extension PCR reactions\cite{yazdi2015rewritable1}. The design of DNA PCR primers (addresses) from a coding-theoretic perspective, which was initiated with~\cite{yazdi2018mutually}, also played a crucial role in scaling up this approach for larger file sizes. 
%Using the approach first described in~\cite{yazdi2018mutually} and extended in~\cite{levy2018mutually}, Microsoft further demonstrated the effectiveness and scalability of PCR-based random access on file sizes of sizes approximately $200$ MB. 
The second innovation was the design of portable DNA-based data storage platforms that utilize nanopore readouts and are accompanied by specialized pilot sequencing, multiple sequencing alignment, and homopolymer coding approaches~\cite{yazdi2017portable}. This development has given rise to new challenges such as \emph{coded trace reconstruction}~\cite{cheraghchi2020coded} and synchronization error-correction. The third milestone involved an expansion of the molecular alphabet to include modified DNA bases~\cite{tabatabaei2022expanding} that can be read using commercially available nanopore devices coupled with deep learning solutions for base classification. Simultaneously, theoretical models for DNA ``storage channels'' have been proposed to rigorously analyze the above-described architectures~\cite{kiah2016codes}, involving overlapping DNA oligos akin to those used in the original storage architectures of~\cite{church2012next,goldman2013towards}, and nonoverlapping information-bearing blocks which model pools without address sequences (see~\cite{heckel2017fundamental,lenz2021capacity} and references therein). These models have been the basis for further research on the fundamental aspects and capacity of DNA storage channels.

Despite the early success in developing DNA-based data storage systems, many issues remain unresolved, with the most important one being the high error rates resulting from associated synthesis, access, and readout processes~\cite{friedberg2003dna,ross2013characterizing,mikheyev2014first} and the extremely high cost of DNA synthesis. Additionally, known designs still lack the computational capabilities required to support operations on data stored in molecular media. 

Sequencing errors have largely been mitigated through the use of existing~\cite{kosuri2014large,tian2009advancing,ma2012dna,ma2012error} and the design of specialized coding approaches~\cite{milenkovic2006design,grass2015robust,kiah2016codes,gabrys2017asymmetric1,lenz2021codes}. These approaches can handle missing oligos, in-oligo sequencing errors, and asymmetric errors caused by the specific molecular topologies of DNA bases. Handling other types of errors in molecular storage has been made possible through a combination of constrained coding, which avoids DNA patterns prone to synthesis or sequencing errors, prefix synchronized coding, which allows accurate access to blocks of DNA without disturbing other blocks in the DNA pool, and low-density parity-check (LDPC) coding~\cite{gallager1962low}, which provide redundancy for combating substitution errors.

A satisfactory solution to the problem of high-cost synthesis is still missing since synthesis is a sequential process that can currently only be made faster through parallel synthesis of shorter sequence blocks and subsequent ligation/attachment (an approach used by \emph{Catalog}, www.catalogdna.com). The \emph{DNA Punchcard} paradigm was introduced in~\cite{tabatabaei2020dna} as a partial solution to the problem of sequential DNA synthesis and joint storage and computing. In this paradigm, native DNA (i.e., DNA retrieved from common bacterial species such as \emph{E. coli}) is used as the storage media. Binary or ternary information is imprinted on the DNA by creating controlled nicks (i.e., holes) at specific locations of the sugar-phosphate backbone. Since the sequence content of the DNA strings is known, retrieving the stored information is straightforward and close to error-free. It involves denaturing the DNA double helix by heating it to temperatures at which the two strands separate, and then aligning the resulting fragments with the known reference. 

Although the DNA Punchcard system experiences a moderate density loss compared to traditional sequence encodings, it offers highly efficient parallel writing and unique massively parallel in-memory computing features~\cite{chen2021parallel,wang2022parallel}. The in-memory computing model, known as SIMDNA, relies on carefully shifting and recreating nicks in multiple information-bearing DNA registers using \emph{strand displacement} reactions and combinatorial design rules. SIMDNA's most appealing feature is its ability to use the same instruction DNA strands to update all registers, regardless of their content.

Additionally, nicks can be overlaid on DNA strands that carry information to include rewritable data, such as metadata. This rewriting process involves sealing the nicks using native ligases~\cite{pan2022rewritable} and then repunching the helix. The Punchcard method and its recent extension, known as DNA Typewriters~\cite{minton2022dna}, which operates \emph{in vivo}, present new challenges related to constrained coding and error correction due to their unique information storage approach. In Punchcard systems, for instance, it is possible to choose nick locations that have significant differences in sequence content to avoid errors during punching. Nonetheless, the placement of nicks must still satisfy certain requirements regarding their distribution on the two DNA strands. Essentially, the placement of nicks should ensure the overall stability of the double-helix complex. These constraints lead to a new coding paradigm for sets and introduce intriguing questions related to \emph{set discrepancy} analysis~\cite{gabrys2020set}.

The aim of this overview article is to provide an accessible introduction to the key components of DNA-based storage systems. These components include DNA synthesis, PCR protocols for random access, synthetic biology concepts like gene editing using CRISPR complexes, sequencing techniques such as shotgun, nanopore, and PacBio sequencing, as well as strand displacement molecular computation paradigms.

Additionally, the article will describe the fundamental concepts behind current DNA-based data storage architectures. It will explore how real biological challenges have influenced the design of coding solutions that are necessary to ensure reliable scaling and operation of these systems. Moreover, it will highlight the importance of expertise in coding theory to inspire new system designs and tackle practical challenges in system implementation. Special attention will be given to reviewing the recent contribution to the field made by the author and her collaborators. 

The manuscript is organized as follows. Section~\ref{sec:bio-prelim} contains a gentle introduction to relevant terminology from synthetic biology and a review of basic properties of DNA molecules. Section~\ref{sec:platforms} describes a collection of conceptually different approaches to DNA-based data storage system design and provides a short review of DNA strand displacement computational paradigms. Section~\ref{sec:coding} presents a review of coding-theoretic results that were specially developed to deal with reliability and implementation issues encountered in DNA-based data storage systems. %Open problems are described in Section~\ref{sec:conclusions}.

\section{Synthetic Biology Preliminaries} \label{sec:bio-prelim}

Deoxyribonucleic acid (DNA) is a macromolecule, which is a molecule made up of a large number of atoms. It is found in the cell nucleus of higher organisms (eukaryotes), which is a compartment in the cell with a width of $5-10\mu m$.

\begin{figure}[h]
\centering
% \begin{wrapfigure}{l}{0.47\textwidth}
% \vspace{-0.15in}
\includegraphics[width=0.6\textwidth]{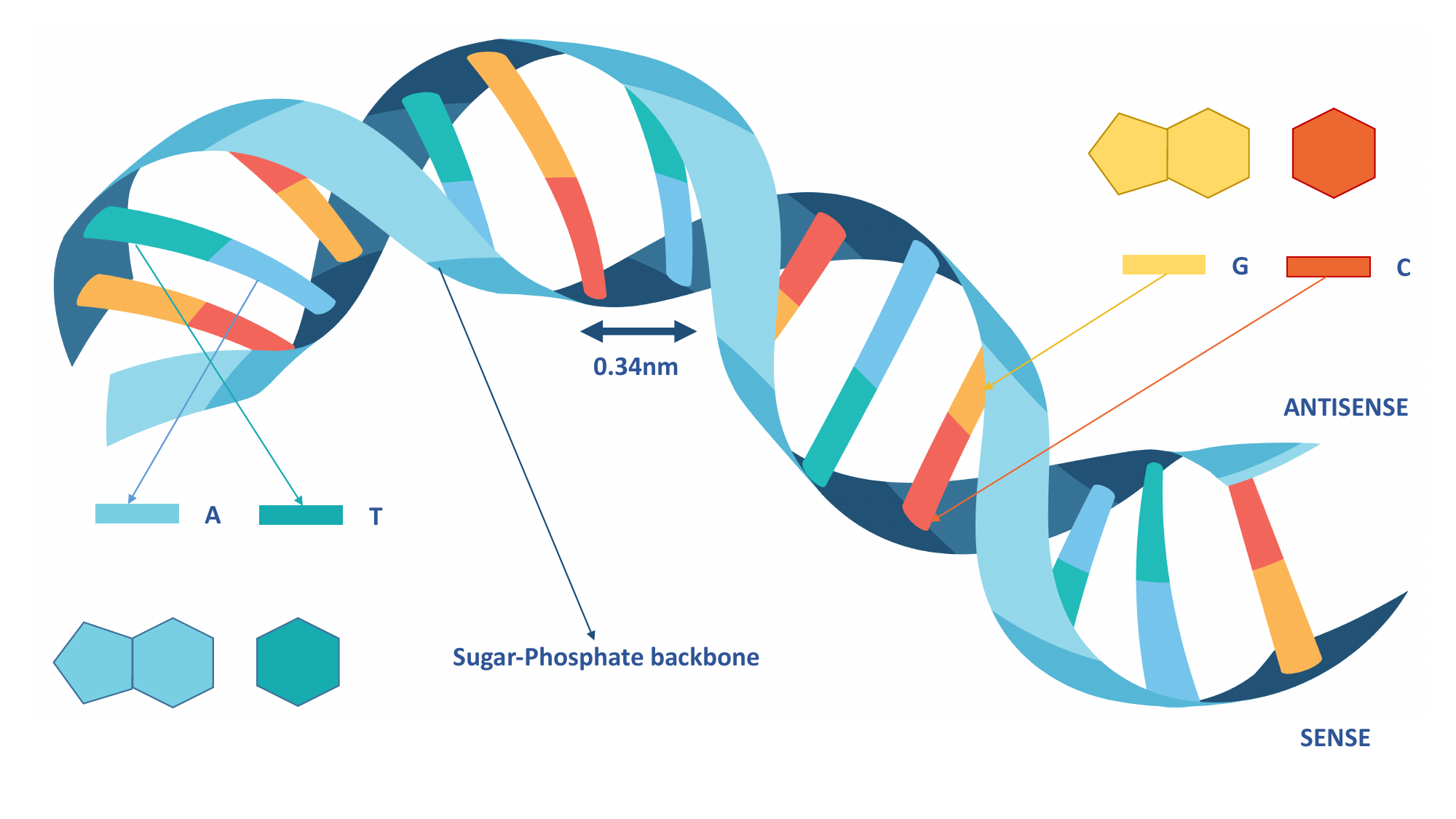}
\vspace{-0.2in}
\caption{Structure of the DNA macromolecule.} \label{fig:dna}
% \vspace{-0.2in}
% \end{wrapfigure}
\end{figure}

In eukaryotes, DNA takes the form of a right-handed double-helix. It consists of two linear molecules that twist around each other, forming the \emph{sugar-phosphate backbone} (see Figure~\ref{fig:dna}). The sugar-phosphate backbone has a deterministic structure, alternating between a deoxyribose sugar molecule and a phosphate group. It does not carry useful information. Useful information is contained in the space between the linear molecules, where four different molecular structures, called \emph{bases}, bind together in pairs through hydrogen bonds. The bases are adenine ($A$), guanine ($G$), cytosine ($C$), and thymine ($T$). $A$ and $G$, which have two carbon rings, are \emph{purine bases}. $C$ and $T$, which have one carbon ring, are \emph{pyrimidine bases}. A \emph{nucleotide} consists of one base, one sugar, and one phosphate group.

DNA of eukaryotes is of the form of a \emph{right-handed double-helix}, with two periodic linear molecules that twist around each other forming the \emph{sugar-phosphate backbone} (see Figure~\ref{fig:dna}). The sugar-phosphate backbone has a deterministic structure, as it alternates between a deoxyribose sugar molecule and a phosphate group. It therefore does not carry useful information. The information-bearing content resides in the space between the linear molecules: four different molecular structures, termed \emph{bases}, including adenine ($A$), 
guanine ($G$), cytosine ($C$) and thymine ($T$), form molecular pairs that bind through \emph{hydrogen bonds}. The bases $A$ and $G$ contain two \emph{carbon rings} (depicted in Figure~\ref{fig:dna} by a hexagon-pentagon structure) and belong to the class of \emph{purine bases}, while the bases $T$ and $C$ contain only one carbon ring (depicted in Figure~\ref{fig:dna} by a hexagon) and belong to the class of \emph{pyrimidine bases}. A molecular unit comprising one base, one sugar and one phosphate group is termed a \emph{nucleotide}, and is often used interchangeably with the term base.

There are two important observations to make about DNA bases. First, not all pairings are possible. According to the \emph{Watson-Crick rule}, $A$ only binds with $T$ through two hydrogen bonds, and vice versa. $G$ only binds with $C$ through three hydrogen bonds, and vice versa. While there are some rare exceptions, the Watson-Crick rule is generally considered a fundamental constraint for DNA molecules. As a result, the information-bearing sequence attached to one linear molecule is the Watson-Crick complement of the information-bearing sequence attached to the other linear molecule. For example, the Watson-Crick complement of $ATTCG$ is $TAAGC$.

Second, since bases are asymmetric molecules, we can orient DNA strings based on the numbering of the terminal carbon atom at the end of the string. Only the $3$rd and $5$th carbon can appear at the terminus of the deoxyribose sugar ring, and a string can be read from either the $3'$ carbon end or the $5'$ carbon end. The symbol $'$ is used to denote the carbon atoms in the sugar ring but should not confuse the reader as it is just part of the chemistry vocabulary. For example, the string $ATTCG$ used in the previous example may be read from the $3'$ to $5'$ end, written as $3'-ATTCG-5'$. This string is different from its reversal, which is written as $5'-ATTCG-3'$. If both base strings of a DNA molecule are read in the same direction, they represent reverse Watson-Crick complements. For the running example, if both strings are read from the $3'$ to $5'$ end, they would be equal to $ATTCG$ and $CGAAT$. Alternatively, they can be written as $3'-ATTCG-5'$ and $5'-TAAGC-3'$. The strand running in the $5'-3'$ direction is called the \emph{sense strand,} and the string running in the $3'-5'$ direction is called the \emph{antisense strand.} These terms are borrowed from genetics and are based on the reading directions of protein-coding genes. Here, they are only used to refer to the orientation of the strands since no genes are involved. Lastly, the process of binding a sense and antisense strand to form a double-helix is called \emph{hybridization,} and the process of separating the sense and antisense strand is called \emph{denaturation.} Denaturation is typically achieved by heating up the DNA, as thermal energy breaks down the hydrogen bonds and leads to the disassociation of the strands.

While not immediately evident, the aforementioned properties of DNA molecules bear significant relevance in the context of DNA-based data storage system implementations.

For instance, consider the two purine bases, $A$ and $G,$ which each possesses two carbon rings, resulting in a more similar chemical structure when compared to pyrimidines. This similarity implies a higher likelihood of confusing them during sequencing, in contrast to, for instance, $A$ and $T$. This observation also extends to the pyrimidines. To address this issue of higher confusability between the pairs of pyrimidine and purine bases and the relatively lower confusability between purines and pyrimidines, specialized \emph{asymmetric Lee distance codes} were proposed in~\cite{gabrys2017asymmetric1}.

Furthermore, the disparity in the number of hydrogen bonds formed between $A$ and $T$ versus $G$ and $C$ in the Watson-Crick pairings underscores the necessity of maintaining what is known as ``balanced $GC$ content'' in information-bearing DNA strings. A small number of $GC$ pairs may lead to instability in the DNA duplex while a large number may hinder efficient DNA synthesis, as elaborated in the following section.

\begin{figure}
\centering
% \begin{wrapfigure}{r}{0.65\textwidth}
% \vspace{-0.1in}
\includegraphics[width=0.75\textwidth]{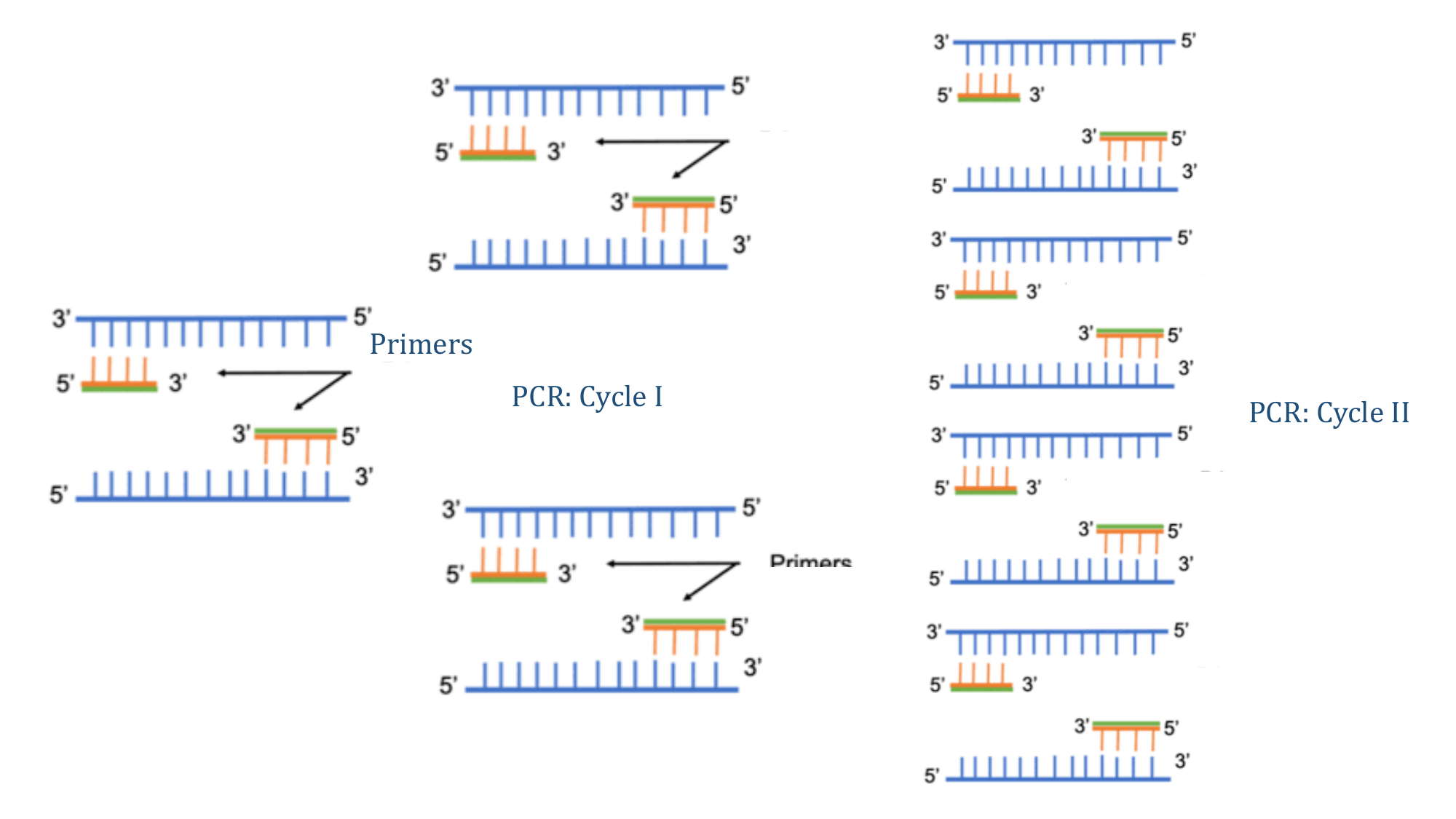}
% \vspace{-0.2in}
\caption{Illustration of primer binding on denatured DNA. Two primers are required, one for the sense strand and another for the antisense strand. In each cycle of PCR amplification, utilizing Watson-Crick complementarity, two identical copies of the same strand are generated, ideally resulting in exponential growth in the concentration of the DNA product. The first random access protocol, described in~\cite{yazdi2015rewritable1}, relies on the use of primers to amplify only a desired collection of DNA sequences. After performing PCR amplification on these DNA sequences for a sufficient number of cycles, there is an overwhelming probability that only the intended sequences are present in the pool. This makes them easily sequenced afterwards. For information on microfluidic and self-rolled membrane random access systems that do not require PCR amplification, the interested reader is referred to~\cite{khandelwal2022self}.} \label{fig:primers}
\vspace{-0.2in}
% \end{wrapfigure}
\end{figure}

The significance of the reverse Watson-Crick strings becomes evident in the context of DNA replication, a process that involves creating two copies of DNA from a single template. During replication, the double helix gradually unravels, allowing each constituent string to serve as a template for generating a complementary strand. Outside the cell, replication is performed through a process called Polymerase Chain Reaction (PCR), which is also employed in the testing of viral diseases like Covid-19 (see Figure~\ref{fig:primers}) and plays a crucial role in the unique approach to random access in DNA-based data storage~\cite{yazdi2015rewritable1}.

DNA replication cannot commence without a specialized class of molecules known as ``primers.'' Primers are short DNA fragments, roughly $20$ bases in length, which are single-stranded. Primers facilitate the binding of enzymes (functional proteins) essential for the replication of the DNA strands. To enable DNA content amplification and, consequently, random access, primers must adhere to several constraints. First, their ``melting temperature,'' defined as the temperature at which $50$\% of the DNA exists in a double-stranded form and $50$\% in a single-stranded form within a solution, must closely match the range of temperatures $55-70^{\circ}$C. Maintaining an appropriate melting temperature and binding stability hinges on a constant, balanced $GC$ content. Numerous online platforms, such as \emph{In silico}\footnote{\url{http://insilico.ehu.es/tm.php?formula=basic}.}, provide precise estimates of melting temperatures.

Second, it is crucial to meet the ``no-folding'' and ``no primer-dimer'' constraints. Folding refers to the formation of a partially hybridized molecule through pairings of complementary bases on the same strand. Primer-dimer constraints, on the other hand, prevent two distinct single-stranded primers from hybridizing or partially hybridizing with each other. For a comprehensive exploration of the coding challenges associated with primer design constraints, interested readers are directed to~\cite{yazdi2018mutually}.
 
The linear distance between two adjacent bases on DNA strings is approximately $0.34$ nanometers, which implies that DNA can store $2$ bits within this length. Consequently, the linear storage density of DNA is approximately $6 \times 10^9$ bits/m. More commonly, storage density is expressed in terms of bit-mass density, taking into account that the mass of a nucleotide is $330$ Daltons, with one Dalton equal to $1.66 \times 10^{-24}$ grams. This translates to the ability to store $2$ bits in a mass of $5.28 \times10^{-22}$ grams, or $3.78  \times 10^{21}$ bits/gram. It is important to note that this represents the \emph{physical storage density}, which is typically higher than the information storage density due to the latter taking into account overheads for address, error-correction, and constraint coding.
 
The reported information densities currently surpass by orders of magnitude those achievable by any other existing storage technology. Furthermore, DNA can maintain its integrity for tens to hundreds of thousands of years when stored in a low-humidity, radiation-free environment. Given the ongoing drive for performance enhancements and cost reductions in DNA writing (synthesis) and reading (sequencing) technologies, especially in the fields of medical and fundamental molecular biology research, molecular storage platforms hold a unique position among their competitors, with minimal concerns regarding future system compatibility.

\subsection{DNA Synthesis and Sequencing: Building a DNA-Based Data Storage System}

Building a basic DNA-based data storage system is indeed feasible, but it entails several essential components: ample financial resources, reliable synthetic DNA suppliers, and access to sequencing platforms such as Illumina or third-generation alternatives including Oxford Nanotechnologies (ONT) or Pacific Biosciences (PacBio). Such sequencers are readily used in genomic research laboratories, but with the exception of ONT systems, they are too expensive and bulky to be part of commercial readout systems.

The availability of ``sufficient funding'' is paramount, given that DNA synthesis commands a substantial cost. This financial requirement stands as the primary impediment to the widespread adoption of molecular storage systems on a large scale. In the ensuing discussion, we will elucidate the principles underpinning DNA synthesis and sequencing while shedding light on potential errors that can arise during these intricate processes.

\textbf{Synthesis.} To synthesize user information into DNA, the initial step involves introducing controlled redundancy into the original binary data string. This redundancy serves two crucial purposes: facilitating various functionalities and ensuring robustness. Subsequently, the original binary information string is transformed into a string over the DNA alphabet comprising four letters, $A,T,G,C$. 
It is worth noting that advancements in chemically modified DNA-based data storage have paved the way for conversions into larger alphabets, spanning from $8$ to $11$ letters~\cite{tabatabaei2022expanding}.

In the next stage, the quaternary data string is segmented into either overlapping or nonoverlapping substrings. These substrings are converted into actual DNA strings harboring identical content. While overlapping substrings were initially employed in the early prototypes of DNA-based data storage~\cite{church2012next,goldman2013towards}, this approach has been mostly abandoned due to its high coding redundancy and inefficiency of random access. When synthesizing the content, it is important to consider two key factors: the lengths of the substrings and the format in which they are delivered, which is constrained by the synthesis technology used.

For instance, when procuring products from \emph{Integrated DNA Technologies (IDT),} customers can opt for what are referred to as \emph{gBlocks.} gBlocks are double-stranded DNA strings with lengths of approximately up to $3,000$ base pairs (bps). They are primarily used for gene construction and play a pivotal role in genome editing\footnote{see \url{https://www.idtdna.com/pages/products/genes-and-gene-fragments/double-stranded-dna-fragments/}.}. Each gBlock is provided as an individual string, and users have the flexibility to choose the molar concentration. Typically, gBlocks require the inclusion of prefix and suffix primer sequences to enable subsequent amplification of the relatively small volume of purchased synthetic DNA. The same primers are used in PCR-based random access. The advantages of gBlocks include their long length, which ensures a smaller proportion of the content dedicated to primer substrings, stable double-helix structure, as well as their ease of reading via ONT and PacBio devices. Additionally, each fragment is provided in a separate storage tube or well. However, it is important to note that gBlocks are associated with a higher synthesis cost per nucleotide when compared to their shorter single-stranded counterparts.

As an alternative, one can opt for ``DNA oligo pools''\footnote{\url{https://www.idtdna.com/pages/products/custom-dna-rna/}.}. These pools consist of unordered collections of numerous short single-stranded DNA strings, referred to as \emph{oligos}. For instance, \emph{IDT oPools} are available in formats that encompass anywhere from $2$ to $384$ oligos per pool, with oligo lengths ranging from $4$ to $350$ nucleotides. It is guaranteed that each oligo is present at a concentration of $50$ pmols. The most cost-effective package offers oPools with a per-base cost of approximately $0.011$. This cost, while significantly higher than that of traditional recording media, still represents a more budget-friendly alternative compared to the per-base cost of gBlocks. 

Notably, oPools come with their own set of advantages and disadvantages. Advantages of oPools include the previously mentioned cost-effectiveness and ease of handling. However, they also exhibit several drawbacks. Typically, oPools have a lower average synthesis fidelity, reduced stability, and a propensity to hybridize with each other. Additionally, they are burdened by substantial primer overheads. Furthermore, if not synthesized to full lengths ranging from $150$ to $300$ bases, they cannot be directly read using third-generation sequencing devices. Detrimental for the underlying molecular storage systems is the \emph{missing oligo} problem, referring to the absence of one or more oligos requested for synthesis. Missing oligo errors arise due to many factors such as the placement of the oligo to be synthesized on a microarray grid, their base content and others. It is also worth noting that the primers required for content amplification need to be purchased separately\footnote{\url{https://www.idtdna.com/pages/products/custom-dna-rna/dna-oligos/custom-dna-oligos}.}.

\begin{figure}[h]
\centering
% \begin{wrapfigure}{l}{0.6\textwidth}
% \vspace{-0.25in}
\includegraphics[width=0.8\textwidth]{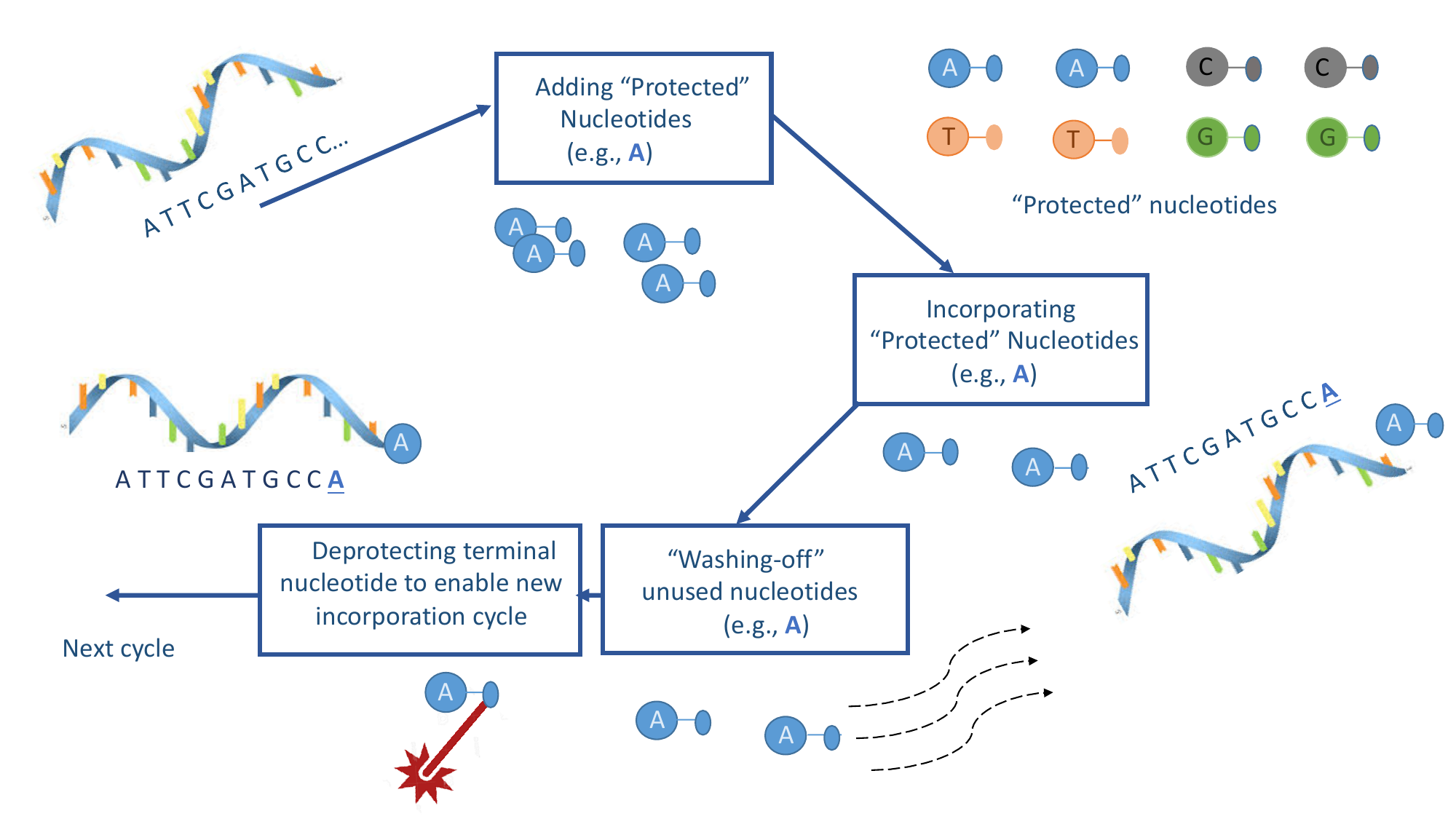}
% \vspace{-0.23in}
\caption{High-level illustration of the synthesis process: one may think of heaving four types of different beads (nucleotides) which have to be stitched together sequentially to create the desired user content. The process is inherently sequential and involves multiple steps of nucleotide incorporation, washing and protective group deactivation that can lead to long synthesis latency (most providers report times of $1-10$ seconds/nucleotide).} \label{fig:synthesis}
% \vspace{-0.3in}
% \end{wrapfigure}
\end{figure}

A simplified diagram of the steps used in commercial DNA synthesis is depicted in Figure~\ref{fig:synthesis}. Specialized forms of all four types of nucleotides that can be attached to a growing DNA strand are kept in four 
separate repositories and retrieved according to the string being synthesized. The nucleotides contain special \emph{protective groups}, depicted as triangles.

Assume, for example, that the string $ATTCGATGCC$ has already been synthesized and that we want to add the symbol $A$ next. In this case, we would flush the synthesis well containing the partially synthesized string with protected $A$ nucleotides and the enzymes (including polymerases) necessary for synthesis. The protective group prevents unintentional incorporation of multiple DNA symbols in one round/cycle of synthesis. Specifically, it disables access to other nucleotides on the strand once it is added as part of the newly included nucleotide. After the nucleotide is incorporated, any unused $A$ symbols need to be washed off to avoid contaminating the new pool of symbols (which may be different from $A$) in the next cycle of extension. Washing is not entirely precise, so some unused nucleotides from previous cycles may remain. However, due to extensive chemical error-correction of the strands, this is not a very likely error event (i.e., in practice, no errors involving repeated symbols are observed in gBlock DNA products, and only a small fraction of errors are typically observed in sequenced oPools, where the errors may have actually been introduced during sequencing). Nonetheless, in theory, simultaneous incorporation of multiple bases could lead to \emph{sticky insertion errors}~\cite{mitzenmacher2008capacity}. Once the washing process is completed, and in preparation for the next cycle of growth for the extended strand $ATTCGATGCCA$, the protective group is removed or deactivated using lasers or other methods. However, the deactivation process is also prone to errors, which may result in some strands being permanently ``deactivated''. In this case, one ends up with incompletely synthesized DNA oligos, which are usually removed by the vendor before delivering the product. In some cases, temporary deactivation leads to oligos with burst deletions. Oligos with bursty deletions can be identified through their shorter length and removed. IDT products have a very low likelihood of containing synthesis errors of the aforementioned type, but the company may report synthesis issues related to unbalanced $GC$-content and \emph{short repeats}, i.e., repeats of short DNA substrings. The most significant errors are typically missing or low-coverage oligos, where certain fragments were not synthesized to acceptable lengths or were synthesized inefficiently. These errors can be corrected using Reed-Solomon coding schemes as described in~\cite{grass2015robust} or through machine learning techniques used in 2DDNA systems~\cite{pan2022rewritable}. Lastly, it is worth mentioning that questions regarding how to improve the efficiency of synthesis schedules were addressed in~\cite{jain2020coding,lenz2021codes,makarychev2022batch}.

\begin{figure}[h]
\centering
% \begin{wrapfigure}{r}{0.55\textwidth}
% \vspace{-0.3in}
\includegraphics[width=0.8\textwidth]{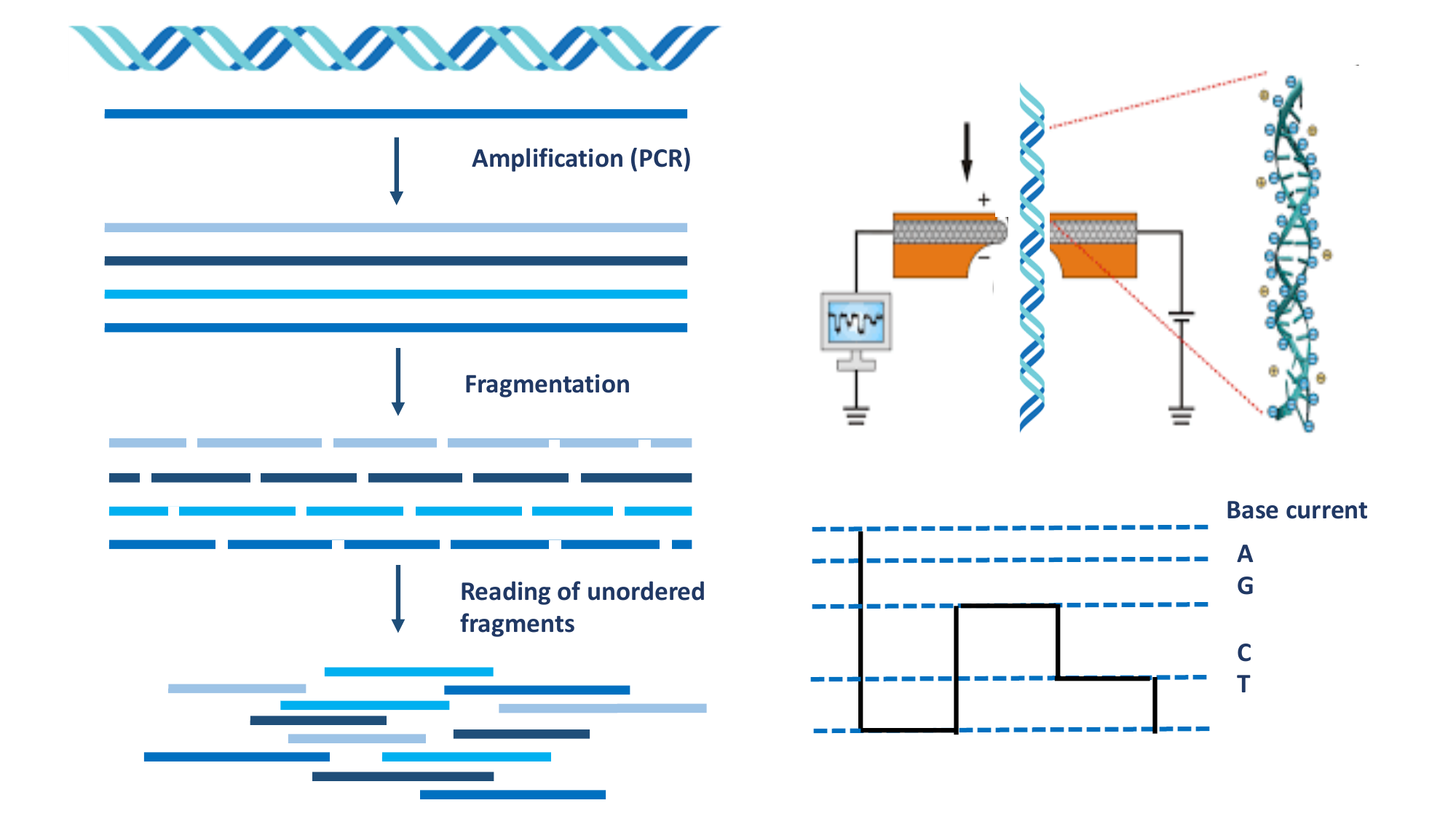}
% \vspace{-0.35in}
\caption{Principles underlying next and third generation sequencing platforms: (left) shotgun sequencing, the idea behind sequencing long DNA strands broken into overlapping fragments that are stitched together during the assembly process. This is also the basic idea of the approach used in Illumina sequencing, along with the unique idea of \emph{bridge amplification}; (right) a fundamentally different approach used in third generation ONT devices, termed nanopore sequencers. There, DNA strands are translocated (passed) through pores (holes), interrupting the flows of ions across the pore. The resulting drop in the ion current is indicative of the charge and structure of nucleotides within the pore.} \label{fig:sequencing}
% \vspace{-0.3in}
% \end{wrapfigure}
\end{figure}

\textbf{Sequencing.} There are currently many different technologies available for reading the content stored in DNA. One example is the sequencing-by-synthesis approach used by Illumina platforms, as shown on the left of Figure~\ref{fig:sequencing}. Illumina devices, such as Miseq, Novaseq, and Hiseq, have a limitation in terms of the length of strands they can read, which is usually not more than $400$ nucleotides. This technology is commonly used for reading pools of DNA oligos because the oligo lengths match the required sequencing lengths. The DNA fragments generated by Illumina and other sequencers are referred to as \emph{reads} and are summarized in raw data files with the .fast or .fastq extension. The .fastq files not only contain read sequences but also information about quality scores of the symbols, allowing for assessment of the quality of the results. Illumina systems have high sequencing accuracy, although still not accurate enough for demanding storage applications (most systems currently operate with an error rate of less than $0.1-1\%$). Additionally, since multiple copies of the DNA strands are read simultaneously, consensus sequences can be easily formed by using majority counts for each position in the reads, as errors are mainly substitution errors.

On the other hand, third generation sequencers are capable of handling long gBlock data formats and are commonly known as \emph{long read technologies.} One important long-read sequencing technology utilizes nanopores, which can provide single-molecule readouts of lengths ranging from $15,000$ to $20,000$ bases, or longer. Nanopores are pores or holes embedded in membranes, with one or multiple pores on the same membrane. In the case of ONT nanopores, the pores are ``biological'' pores, such as proteins, and double-stranded DNA is used for sequencing to control the speed at which it translocates through the pore. This control is achieved through biological motors, often helicases, which unwind the DNA and slow down the passage of one of the strands through the pore. Consequently, ONT systems can only sequence double-stranded DNA.

By applying a voltage current across the membrane, an ion current is maintained within the pore. In the absence of any molecules to be sensed, this current is referred to as the \emph{base current.} When single-stranded DNA translocates through the pore, the short DNA fragment (approximately $3-5$ bases in length, known as a $k$-mer, with $k=3,4,5$) that fits into the pore causes a drop in the ion current as it blocks the movement of ions. As the DNA is moved one base at a time, the duration in which the specific $3/4/5$-mer is observed and the current is recorded should be sufficient to estimate the sequenced DNA. Additionally, the drop in current depends on the $A, T, G, C$  content of the sequenced DNA. Generally, the current drop is influenced by the charge, 3D structure/shape of the nucleotides, and other factors.

Similar to other sequencing technologies, each DNA fragment is replicated before sequencing to obtain multiple reads for reconstructing the original content. The reads corresponding to the same information string are passed through different pores and/or at different times through the same pore. As a result, the reads may exhibit varying levels of sequencing noise. Typically, the process of deciphering the current readouts using multiple reads, known as ``nanopore base calling,'' is facilitated by deep learning approaches involving convolutional and recurrent neural networks (CNNs and RNNs, respectively), which are described in more detail in~\cite{wick2019performance}.  

Based on the previous discussion about the similarity of nucleotide chemical structures and the impact of $k$-mers on the current drop, it is evident that the accuracy of base calling in nanopores is lower compared to Illumina platforms. However, recent reports from ONT indicate significant progress in improving read reliability. According to reports for R10.4 sequencing flowcells, the error rate for single molecule consensus is estimated at $>0.1\%$. In academic labs, the observed error rates appear to be higher than $0.1\%$, with contributions from substitution, deletion, and insertion symbol errors.

The formation of consensus reads in nanopore sequencing is similar to the corresponding process in short-read technologies, but aligning them is computationally more challenging due to the presence of indel errors -- see the description of \emph{multiple sequence alignment algorithms} reported in the context of DNA-based data storage in~\cite{yazdi2017portable}, including Muscle, Coffee, Clustal Omega and others. The work~\cite{yazdi2017portable} also introduced a specialized approach for error-correction from base-called reads using symbol-level redundancy, treating the problem as an instance of \emph{trace reconstruction}. Trace reconstruction was initially described in the context of phylogenetic tree analysis~\cite{batu2004reconstructing}. The reconstruction problem can be summarized as follows: one is given a string over a finite alphabet which is passed through parallel deletion channels, each of which introduces i.i.d deletions. The observed strings at the outputs of the channels are referred to as \emph{traces}. The question of interest is to determine the smallest number of traces needed to reconstruct the original string in the \emph{average and worst-case scenario} for a given deletion probability.

The connection to nanopore sequencing is immediate by associating each trace with a read generated by one or multiple pores. The problem of trace reconstruction in which the DNA strings are allowed to be encoded is known as coded trace reconstruction~\cite{cheraghchi2020coded}, and is discussed in more detail in subsequent sections. Other nanopore error-correcting codes that directly operate on raw current readouts without requiring intermediate basecalling are discussed in~\cite{chandak2020overcoming}. Some other interesting results on reconstructing strings based on traces and modeling the nanopore channel can be found in~\cite{magner2016fundamental,mao2018models,mcbain2022finite}.

Although current DNA-based data storage systems do not effectively utilize PacBio HiFi technologies~\cite{lang2020comparison}, it is important to highlight some notable features of this technology. HiFi sequencers produce long reads, ranging from $10,000$ to $20,000$ bases, and exhibit high reliability comparable to Sanger sequencers. This increased accuracy in base calling can be attributed to various factors, including the reduction of polymerase bleaching effects and the implementation of subread consensus protocols. In the HiFi sequencing process, the same DNA molecule is read approximately $200$ times, generating an equal number of subreads that are subsequently aligned and denoised. Unlike nanopores, HiFi devices capture the \emph{kinetics} of the reading process, where each base is characterized by distinguishable \emph{pulse widths,} and each pair of bases corresponds to different \emph{interpulse widths.} These pulse width and interpulse duration signals reflect the speed at which a polymerase incorporates a specific base into the subread. Our focus in subsequent discussions is exclusively on long-read nanopore-based DNA storage systems.

\subsection{DNA Editing}\label{sec:crispr}
\begin{figure}[h]
\centering
% \begin{wrapfigure}{l}{0.52\textwidth}
\vspace{-0.2in}
\includegraphics[width=0.7\textwidth]{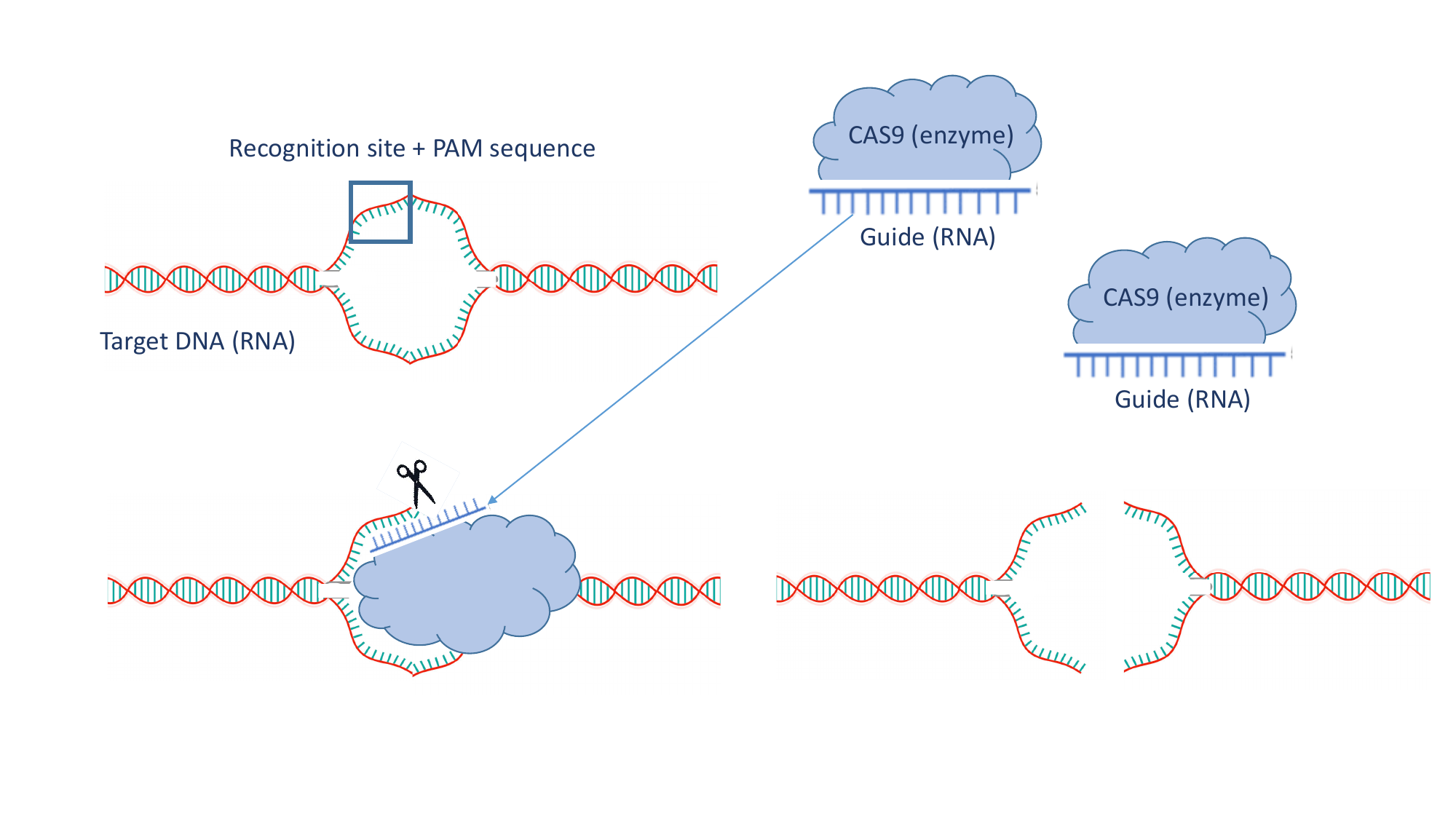}
\vspace{-0.4in}
\caption{The CRISPR system and its constituent Cas9 protein and guide RNA components. CRISPR memorizes genetic information of past invasive species and uses it to identify their renewed presence. Upon detection, it performs cutting of the genetic content of the infective agents in order to disable their replication. Outside of this native context, the complex can be used to cut arbitrary DNA strings through a careful design of the RNA guides.} \label{fig:crispr}
\vspace{-0.1in}
% \end{wrapfigure}
\end{figure}

DNA editing is an emerging interdisciplinary field with applications in chemistry, biology, medical sciences and synthetic biology concerned with alternating the content and structure of genomic and other -omic sequences. One of the major breakthrough discoveries in the area is the \emph{CRISPR} (Clustered Regularly Interspaced Short Palindromic Repeats) system which was recently recognized by a Nobel prize in Chemistry awarded to its co-inventors, Charpentier and Doudna~\cite{jinek2012programmable}. CRISPR is a system native to several archaea and bacteria that use it as a form of immune and antiviral defense mechanism. The system involves repeat sequences of certain genetic sequences interleaved (interspersed) with spacer sequences that represent identifiers of invasive species encountered in the past. Upon detection of a recurrent invading unit through recognition of its characterizing genetic sequence, CRISPRs constituent Cas9 proteins guided by RNA recognition sequences cut the viral genomes at the position of the recognized content. Simply put, CRISPR stores snippets of genetic information of prior invasive species and uses this ``genetic memory'' to detect and disable present hostile viruses by cutting their genetic material (see Figure~\ref{fig:crispr}).

An advantage of CRISPR is that it is a complex that already involves enzymes such as Cas9 and relevant guide RNA sequences needed for disabling invasive species. Also, note that the complex can perform cutting of single-stranded and double-stranded substrates in different manners. When cutting double-stranded DNA, either both strands or only one of the two strands can be cut. The latter process is usually referred to ``nicking,'' and it does not lead to disassociation of the duplex. There are also CRISPR complexes involving other enzymes, such as Cas13, which has the capability to edit RNA sequences. 

For DNA-based data storage applications, and in particular, subsequently discussed DNA Punchcards platforms, Cas9 can be matched with arbitrary synthesized guide RNA strings. These lead the enzymes to selected target positions, so that nicking may be performed in a massively parallel fashion involving multiple DNA sites. This is an especially important feature for molecular storage as it circumvents the problems associated with inherently sequential DNA synthesis: if nicks are to be introduced at multiple sufficiently distant locations in a double-stranded DNA substrate, the Cas9 enzymes can perform information recording without co-interference. One drawback of Cas9 is that it is what is known as a \emph{single turnover} molecule -- once the enzyme creates a nick it becomes inactive. This problem can be resolved by using other enzymes (e.g., Pfago~\cite{tabatabaei2020dna}) that are multiple turnover enzymes that, in principle, can make close to hundreds of cuts or nicks before becoming inactive.

\subsection{Strand Displacement}
\emph{Strand displacement} in DNA is one of the most frequently used molecular and DNA computing paradigms. DNA strand displacement, as its name suggests, corresponds to replacing (part of) a single-stranded DNA section of a double-stranded DNA formation by another strand. There are two approaches to displacement: polymerase-based and \emph{toehold-mediated}. We focus on toehold-mediated strand displacement as it is used more frequently, does not require specialized enzymes and lends itself to a wide variety of computations suitable for data stored in DNA Punchcards~\cite{qian2011scaling}. 

\begin{figure}[h]
\centering
% \begin{wrapfigure}{r}{0.55\textwidth}
% \vspace{-0.4in}
\includegraphics[width=0.75\textwidth]{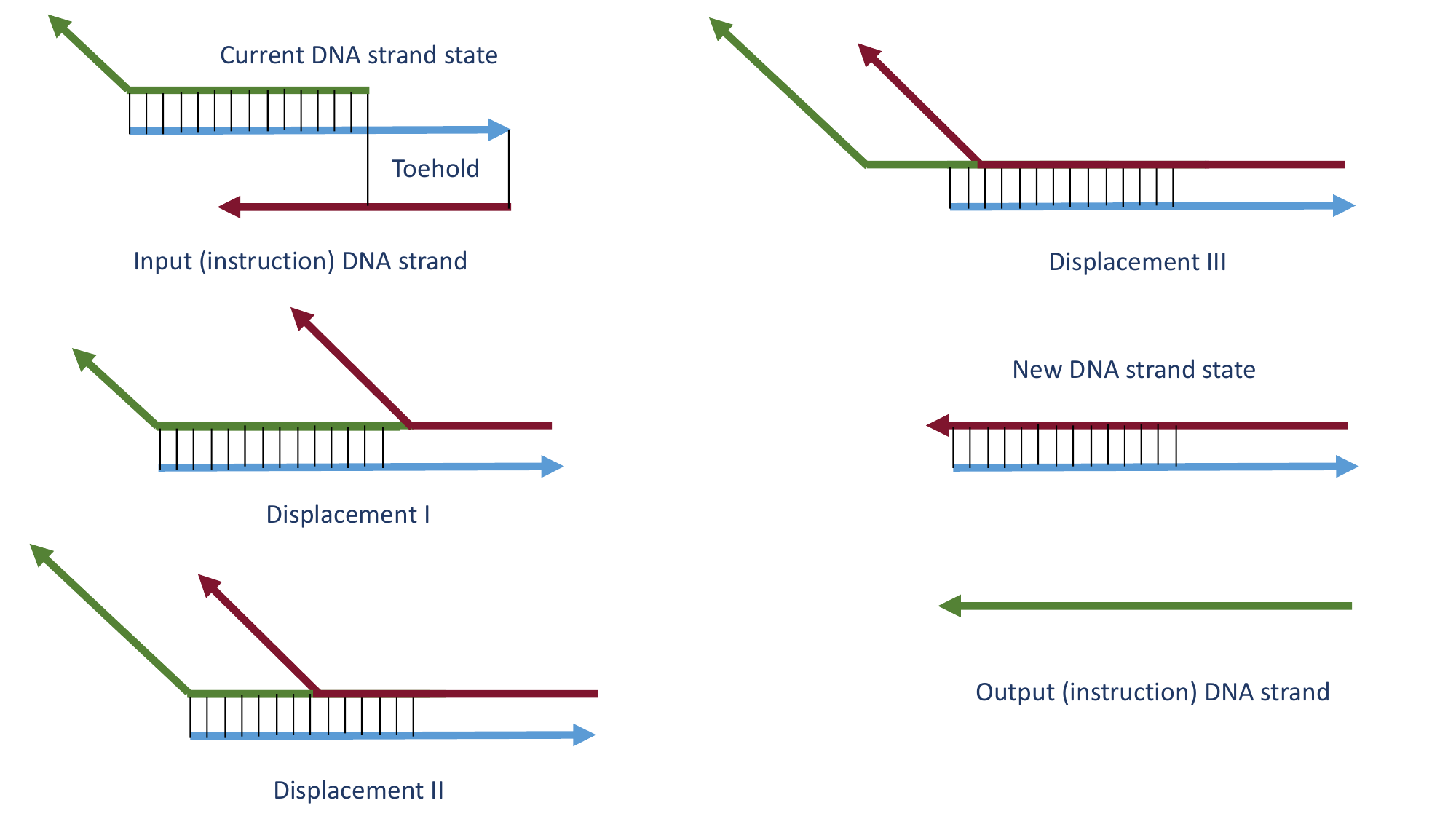}
% \vspace{-0.22in}
\caption{Toehold-mediated strand displacement. The input DNA strand hybridizes to the single-stranded toehold region and forces the competing strand to peel-off from the duplex, as governed by the laws of thermodynamics. The reactions take milliseconds or less.} \label{fig:displacement}
% \vspace{-0.3in}
% \end{wrapfigure}
\end{figure}

The double-stranded DNA may be seen as encoding the state of a computational system or serve as a proxy for a logical gate. It comes with one or multiple single-stranded regions termed \emph{toeholds,} which are usually of length $5-10$ bases. In Figure~\ref{fig:displacement}, there is one toehold in the right-most position of the double-stranded DNA that allows for hybridization of a Watson-Crick-complementary single-stranded DNA, referred to as the \emph{instruction strand}. Once the instruction strand hybridizes to the toehold it starts pushing out (i.e., displacing) the already present single-stranded part of the duplex to the left of the toehold until it completely disassociates. This strand then becomes the ``output'' of the computing unit. In a nutshell, the input strand may be seen as an instruction that changes the state and releases an output strand in its stead. Displacement reactions can be performed in a cascade, thereby allowing for multiple changes of states and released output strands which broadens the computational repertoire of strand displacement. As an example of the computations possible via strand displacement, the interested reader is referred to a neural network implementation based on cascades of toehold-mediated displacements~\cite{qian2011neural}).

Although in theory many different computations, including universal ones, can be implemented via strand displacement, a major practical challenge is to control 
\emph{leakage} in the cascades~\cite{thachuk2015leakless}. Leakage refers to unintended displacements that lead to the release of incorrect output strands and reduce the efficiency of the reactions. Leakage is the key impediment to accurate execution of more than $6-7$ consecutive displacement reactions, due to an excessive number of undesired byproducts. Methods for correcting leakage errors via controlled redundancy were recently described in~\cite{wang2018effective}.

\section{An Overview of Existing DNA-Based Data Storage Platforms} \label{sec:platforms}

\begin{figure}[h]
\centering
% \begin{wrapfigure}{l}{0.5\textwidth}
% \vspace{-0.32in}
\includegraphics[width=0.8\textwidth]{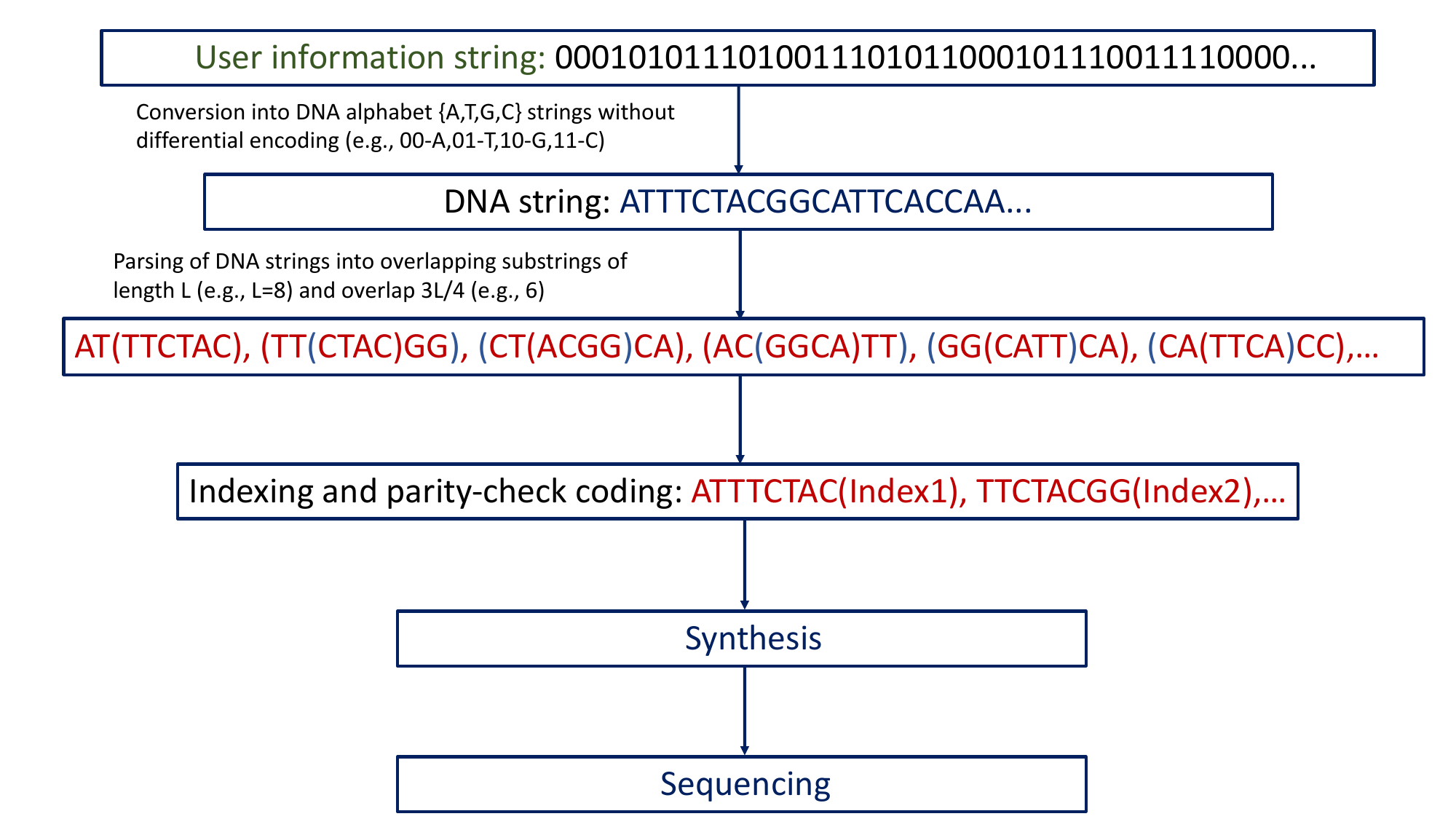}
% \vspace{-0.2in}
\caption{The first two DNA-based data storage systems used nearly identical data encoding protocols, involving runlength coding, indexing and single parity-check coding.} \label{fig:churchgoldman}
% \vspace{-0.3in}
% \end{wrapfigure}
\end{figure}

The first successful implementations of DNA-based storage systems with read and write capabilities were described in ~\cite{church2012next,goldman2013towards}. These works outlined similar procedures, which involved the following steps:
\begin{itemize}
\item[] \textbf{a)} Conversion of compressed binary data, such as text or images, into a ternary or quaternary alphabet. This process included elementary coding techniques like $GC$-balancing, runlength coding, and single-parity check coding. Ternary encoding was employed to limit the runlengths of the same symbol (i.e., homopolymers) to one. It effectively reduced the alphabet size from $3$ to $4$.  
\item[] \textbf{b)} Parsing the encoded information strings into overlapping substrings with controlled overlap length. The overlap was set to $75\%$, ensuring $4$-fold coverage of the content.
\item[] \textbf{c)} Adding substring \emph{identifiers} that encoded the index of each substring within the longer string. Note that such identifiers do not amount to addresses since they were not designed to enable random access.
\item[] \textbf{c)} Synthesizing the overlapping substrings in the form of oligo pools.
\item[] \textbf{d)} Sequencing the substrings and reconstructing the original message (refer to Figure~\ref{fig:churchgoldman}).
\end{itemize}

The system was designed with several main considerations in mind. First, the overlapping oligo approach was designed to facilitate easy reassembly of the original message through identification of overlapping suffixes and prefixes. It effectively reduced the code rate to roughly $1/4$. Second, requiring a balanced $GC$ content was necessary to address both synthesis and sequencing challenges. Third, since sequencing platforms in use at the time of publication, such as Roche $454$, were known to introduce errors in the presence of long homopolymer regions, the latter was severely constrained in length. However, it is important to note that current sequencing platforms do not share homopolymer-related limitations, making it undesirable to compromise storage density through such a restrictive encoding protocol.

The DNA storage systems described were unable to accurately reconstruct the original sequence, despite reducing the size of the alphabet and ensuring long overlaps between adjacent oligos. Furthermore, in order to access the encoded content in a specific section of the sequence, the user had to sequence and assemble the entire content, leading to significant implementation expenses.

The PCR-based random access approach was introduced and experimentally tested as described in~\cite{yazdi2015rewritable1}, with its scalability further confirmed by Microsoft Research on a file size close to $200$ MBs. The idea behind PCR-based random access is simple when analogies to classical storage systems are drawn: one needs to endow each information block (oligo) with an address sequence. The main challenge was to devise a method to efficiently search for the block with the required address when no ``search circuits'' are available. The obvious answer is through hybridization, since the presence of a particular address sequence can be detected via targeted hybridization with its reverse complement sequence. This detection approach is incomplete, since one needs to \emph{isolate} the detected strands and \emph{sequence} them. Isolation is achieved via amplification, i.e., PCR reactions. More precisely, the protocol for RA involves extracting a small subsample of the oligo mixture and running sufficiently many cycles of PCR reactions with primer combinations corresponding to the encoded information blocks to be retrieved. The amplified subsample in this case contains, with overwhelming probability, only the desired oligo content which can be sequenced to complete the access scheme.

The combinatorial design protocol for random access primers includes balancing the $GC$ content, adding error-correcting redundancy, preventing self-folding of the primers, and ensuring that pairs of distinct primers do not hybridize to each (i.e., preventing primer-dimers). Importantly, in addition to all these constraints having to be met simultaneously, one more constraint has to be accounted for -- zero-valued \emph{cross and autocorrelation}~\cite{guibas1981string} of the primers. For details, see~\cite{yazdi2018mutually} and the review article~\cite{yazdi2015dna1}. With regards to questions related to the enumeration of possible single-stranded DNA folds and nonfolding criteria, the interested reader is referred to~\cite{orlitsky1996edge,milenkovic2007constrained}. The latter work used the notion of \emph{Motzkin paths}~\cite{oste2015motzkin}, which may be viewed as Dyck paths augmented with flats. A Dyck path is a string of even length $n$ over the alphabet $(,)$ containing exactly $n/2$ symbols of each type and satisfying the property that no prefix of the string contains more $)$ than $($ symbols. A Motkin path is a string of even length $n$ over the alphabet $(,),-$ containing the same number of $($ and $)$ symbols, and satisfying the property that no prefix of the string contains more $)$ than $($ symbols. An example Dyck path of length $8$ is $(()()())$, while an example Motzkin path of length $8$ is $(-()-()-)$. The matched bracket symbols $($ and $)$ are reserved for paired bases, while the dash $-$ is used to denote unpaired bases. Restricted Motzkin paths described in~\cite{milenkovic2007constrained} ensure that no short collection of consecutive unpaired bases (loop) is followed by a long stem (a pair of reverse-complementary substrings on the string) which would lead to a stable secondary structure.

In addition to reporting the first PCR random access, the work reported in~\cite{yazdi2015rewritable1} also included a text rewriting scheme based on overlap-extension PCR; it also examined, from the theoretical point of view, how to perform information encoding so as to avoid substrings that are identical to the oligo primers used for addressing. In the former setting, a prefix-synchronized coding scheme adapted from~\cite{morita1996construction} was used to ensure that the primer strings do not appear as substrings inside the information-bearing content, as that would lead to PCR amplification of substrands and not the whole oligos. In the context of rewriting, specialized encoding techniques were used to ensure that complete word phrases, likely to be edited together, are part of the same block that can be replaced by another block via so-called overlap-extension PCR reactions. Sequencing was performed using Sanger methods, with no reported errors in the PCR-retrieved information.

Another important direction in DNA-based data storage was taken in~\cite{yazdi2017portable} where for the first time, nanopore sequencing was used for data retrieval instead of standard high-throughput methods. The work also described how to use and combine the ideas of pilot signaling (from communication theory) and trace reconstruction~\cite{batu2004reconstructing} (from theoretical computer science) in DNA-based data storage. At the time the work was published, nanopore sequencers were the only low-cost and portable option for sequencing long blocks of DNA, such as gBlocks. Cost and portability are important issues given that Illumina platforms are bulky and expensive, designed for lab use. Despite their desirable properties, ONT MinION sequencers had the serious drawback of excessively high rates of indel errors, often exceeding $15\%$. 

A common approach to reducing the error rate is to form a consensus of all the nanopore readouts corresponding to different copies of the same input sequence. This naturally leads to the problem of \emph{sequence alignment}, for which software suites such as Clustal Omega~\cite{sievers2014clustal} or ONT in-house learning-based methods such as Nanopolish~\cite{wick2019performance} were and are readily available. Still, for indel error-rates as high as $15\%$, the resulting consensus provided only a low-quality estimate for the actual user string.

\begin{figure}
\centering
% \begin{wrapfigure}{r}{0.54\textwidth}
% \vspace{-0.07in}
\includegraphics[width=0.75\textwidth]{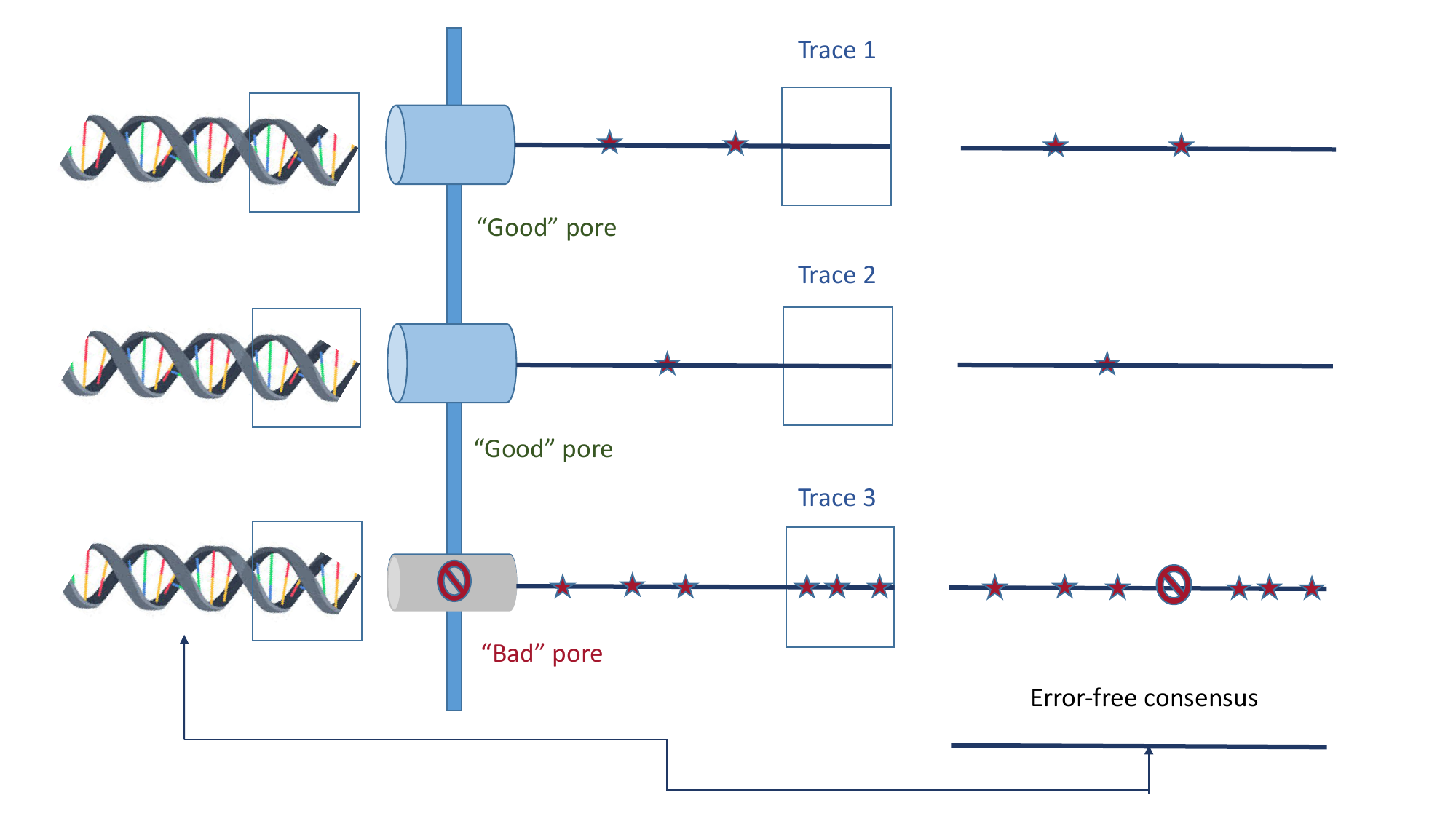}
% \vspace{-0.3in}
\caption{The content of replicas of the same DNA string is read by different pores or by the same pore at different times. Each readout can be modeled as a \emph{trace} (which, unlike most traces used for mathematical proofs of reconstruction performance, also includes insertions and substitution errors). The traces/reads can be aligned using one of the many existing DNA sequence alignment algorithms, but the results are of poor quality when even a small number of traces contains an excessive number of errors (see Extended Data Figure 3 in the Supplementary Information of~\cite{yazdi2017portable}). Since the address sequences of the strings
are known, the quality of the ``nanopore channel'' can be estimated by counting the number of errors present in the address string. In the given example, the address portion of the strings is boxed and the third trace contains three errors in the boxed region. This is indicative of a defective or ``tired'' pore and hence the third trace is not used for sequence alignment. As a result, the consensus sequence obtained via alignment of ``good reads'' is error-free or almost error-free.} \label{fig:traces}
\vspace{-0.2in}
% \end{wrapfigure}
\end{figure}

A straightforward solution to the problem is to treat the addresses as pilot sequences used to estimate the nanopore channel, see Figure~\ref{fig:traces}. This approach proved successful since the addresses/pilots are indicative of either malformed or ``tired'' pores. For such pores,  all traces or the most recently read traces at their outputs, respectively, have a high number of errors. Given that the address sequences are known to both the encoder and decoder, the quality of the pore can be assessed through the number of errors in the addresses. By only using reads whose addresses have no errors, or by iteratively recruiting reads with low-error-rate addresses to improve some local alignments, the reconstruction error rate dropped significantly, below $1-2\%$ for the ONT system used at that time. The remaining errors were completely removed by $GC$-balancing the content of the blocks and by applying asymmetric homopolymer codes. The former allows one to identify potential synchronization or substitution errors by counting the symbols in each subblock; the latter can fix asymmetric deletion errors that affect one or two bases only, and do not completely erase a runlength (homopolymer). Such errors were found to occur in the ONT data generated by the experiments in~\cite{yazdi2017portable}.

Two unconventional approaches to DNA-based data storage were explored in recent studies~\cite{shipman2017crispr,tabatabaei2020dna}. 

In the first approach, synthetic data was incorporated into the DNA of living organisms, such as bacteria. This \emph{in vivo} (inside the cell) approach offers several advantages. First, user-defined information can naturally replicate itself through the growth of bacterial communities. Additionally, this population encoding strategy provides inherent error protection.

However, there are several drawbacks to this scheme. The storage density is relatively low compared to other methods due to the need to carefully place synthetic DNA in specific regions of bacterial genomes, so as not to disrupt normal cellular functions. Furthermore, the ratio between the ``information-bearing mass'' and the overall cell mass is relatively small as well, further decreasing the effective storage density. Most detrimentally, the process of recording and retrieving data is highly complex. Not only does one need to synthesize user information in DNA, but must also insert it into desired locations within the bacterial genome. Data retrieval involves extracting bacterial DNA, isolating the desired content, and subsequently sequencing it. It remains uncertain whether this approach can be made cost-efficient enough to complete with purely synthetic \emph{in vitro} (outside of the cell) methods.

\begin{figure}
\centering
% \begin{wrapfigure}{l}{0.58\textwidth}
% \vspace{-0.3in}
\includegraphics[width=0.8\textwidth]{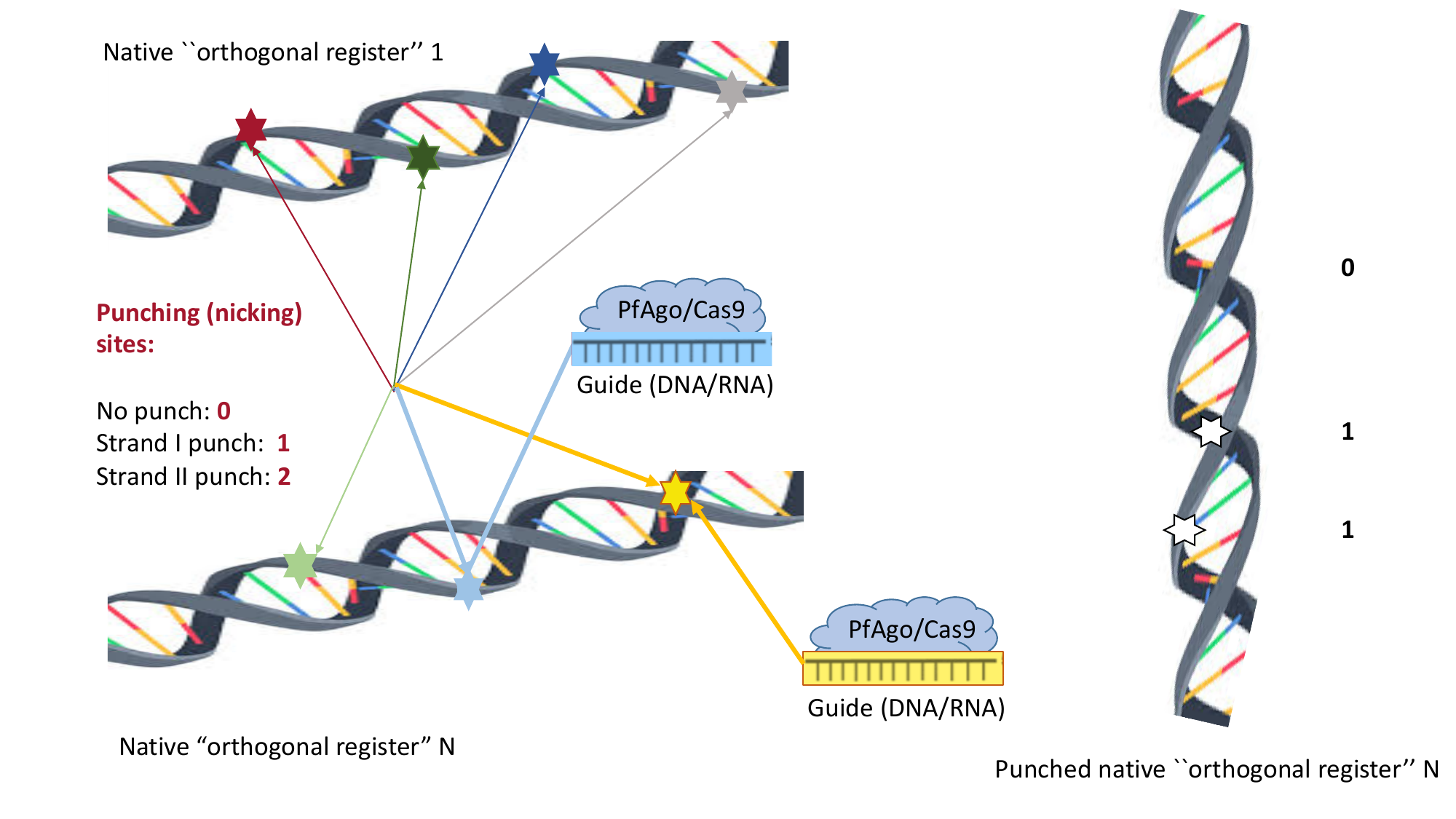}
\vspace{-0.1in}
\caption{In DNA Punchcard systems, data is recorded on native DNA fragments termed orthogonal registers. Each register has nicking locations whose sequence contexts are highly dissimilar, as captured by the different colors of the stars indicating the nicking positions, thereby allowing for parallel nicking/recording of information on all orthogonal strands. The same nicking enzyme-guide unit can be used to nick hundreds of copies of the same register, which makes the system cost-effective.} \label{fig:punch}
\vspace{-0.2in}
% \end{wrapfigure}
\end{figure}

In contrast, the DNA Punchcard system, introduced in~\cite{tabatabaei2020dna}, aims to address the issue of synthesizing DNA in the first place. The concept behind this approach is illustrated in Figure~\ref{fig:punch}. In this storage context, ``native DNA'' refers to DNA extracted directly from bacteria, such as \emph{E. coli}, without any synthetic modifications, and subsequently used and processed  \emph{in vitro}. Native DNA is readily available and can be obtained in large volumes at a low cost. However, since native DNA has a composition determined by nature, it cannot be easily altered to store user-defined data. Instead of modifying the content, one can instead choose to alter the \emph{topology} of the sugar-phosphate backbone at specific positions, known as ``nicking positions.'' These positions are located between a pair of bases and indicate where the backbone strings are allowed to be nicked. Enzymes such as Cas9 or PfAgo, described in Section~\ref{sec:bio-prelim}, can be used for the recording process via nicking. The absence of a nick represents the value \textbf{0}, while a nick on the sense strand represents \textbf{1}, and the same nick on the antisense strand represents \textbf{2}. Therefore, the recording alphabet in this system is ternary. It is important to note that cutting both strands at the same location is not allowed as it would cause the DNA to dissociate.

Storing information through nicking offers several advantages. First, there is no need to synthesize the information content in DNA, as topological changes are utilized to represent the data. Second, nicking can be performed in parallel, making it faster than sequential recording via synthesis. Third, writing the symbol \textbf{0} does not require any specific action, which is a characteristic shared by other existing storage technologies. Fourth, erasing and rewriting data is remarkably straightforward through the use of \emph{ligases,} which can seal off the nicks. Since ligases remove nicks regardless of their position on the DNA, selective erasing and rewriting of data requires storing it in physically separated fragments of DNA. 

The process of reading information stored in nicks is conceptually simple and highly robust to errors due to the existence of the bacterial genome reference. In a nutshell, the nicked DNA is denatured, i.e., the constituent strands are separated, and the obtained fragments PCR-amplified and sequenced using Illumina platforms. The sequenced reads are then aligned to the bacterial reference sequence to determine the locations where one fragment ended and another one started, corresponding to the positions that were actually nicked. Due to the possible presence of short fragments due to closely-placed nicking sites, reading via nanopores would have to be performed using solid-state rather than ONT nanopores, since the former does not unwind the strands~\cite{athreya2020interaction}. 

Another important observation is that it is not necessary to use long native DNA fragments to encode information. 
This is because using long fragments can lead to undesired and off-targeted nicking. Instead, one can selectively isolate 
nonoverlapping fragments of native DNA that have low sequence similarity. These fragments are referred to as ``orthogonal registers.'' By insisting on low sequence similarity, measured in terms of the Levenshtein distance, the probability of nontargeted nicking is reduced while maintaining recording parallelism. Additionally, since the registers are substrings of a real bacterial genome, their order is determined by their occurrence in the genome. Therefore, there is no need for positional encoding. 

Similarly to other current molecular recording systems, DNA Punchcards cannot avoid certain functional impairments. The most important drawback is reduced storage density, which is a consequence of both the decrease of the alphabet size from a $4$-letter base alphabet to a $3$-letter nicking alphabet, as well as the nicking site placement constraint. The latter requires nicking locations to be separated roughly $10$ bases apart in order to ensure the stability of the DNA duplex and avoid having to read very short genomic fragments. 

Note that the guides used in conjunction with the nicking enzymes still need to be synthesized unless they can be extracted directly from the native DNA itself without the use of other guides, which may be challenging. However, guides are typically very short RNA or DNA strings, of length $\leq 20$ nucleotides. Furthermore, the guides are multi-use entities when combined with enzymes like PfAgo. PfAgo has the ability to create hundreds of nicks before becoming inactive.

Nick-based storage allows for in-memory computations to be performed directly and in parallel on the data recorded in all registers through strand displacement operations~\cite{chen2021parallel,wang2022parallel}. In this computing approach, the symbols $0$ and $1$ are stored as two different blocks of bases nicked at different locations. For example, if $5$ nucleotides are used, $0$ could be represented as $2-nick-3$ while $1$ could be represented as $4-nick-1$, indicating that for the former, the nick is placed between the second and third base, while in the latter, it is placed between the fourth and fifth base. Since strand replacement operations terminate when encountering a nick (as the nick prevents further ``peeling-off'' of a DNA substrate) and since nicks encode the symbol values themselves, one can move the positions of the nicks around thereby changing the register content. In a nutshell, these nick-displacement operations involve sealing a nick in one position while creating a new nick in another position. Operations such as incrementing all registers, sorting their contents, and operations behind the universal Rule 110 automata have been successfully implemented and executed on data stored in DNA nicks via multistage strand displacement~\cite{chen2021parallel,wang2022parallel}.

\begin{figure}
% \vspace{-0.15in}
\includegraphics[width=1.0\textwidth]{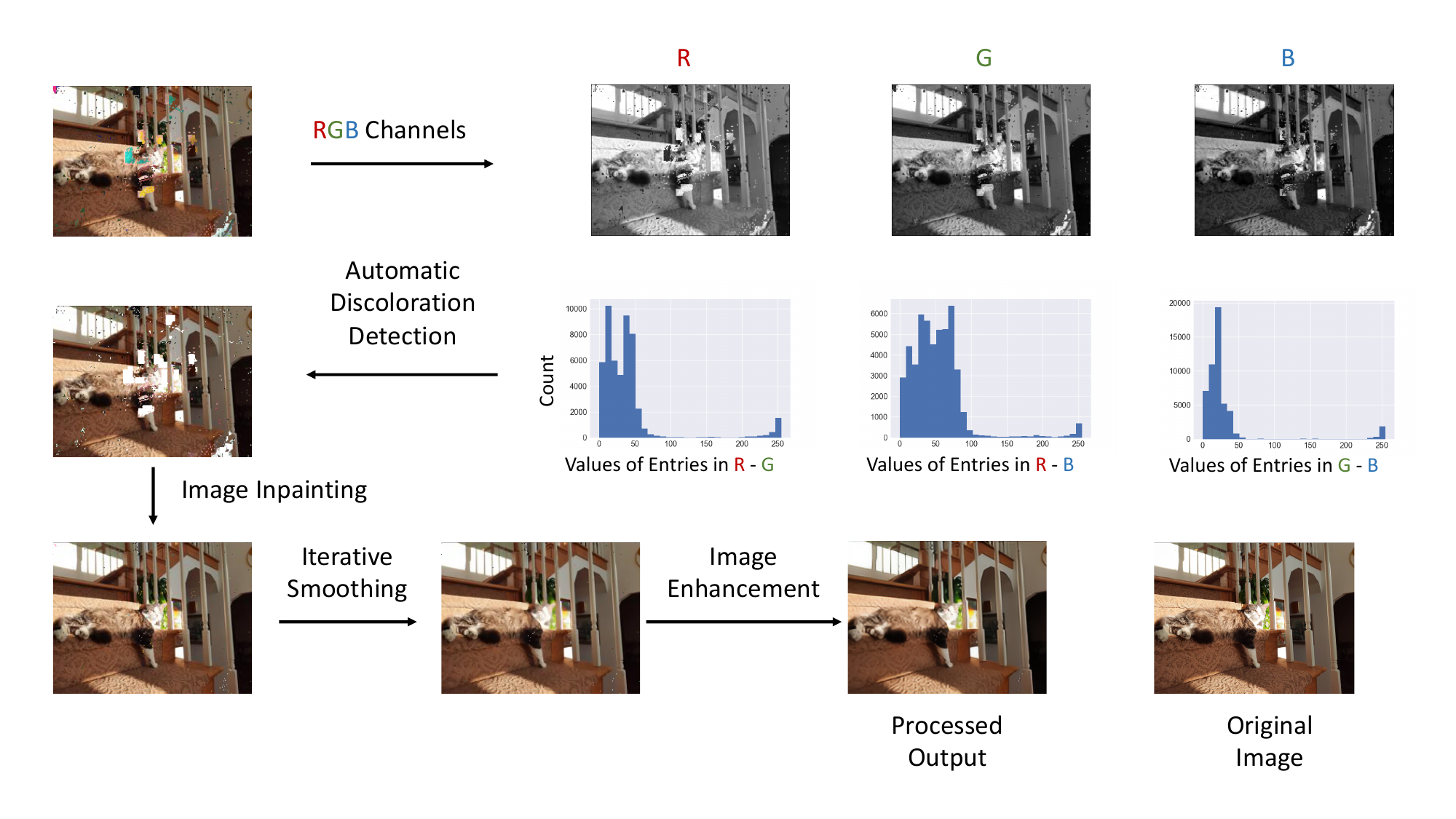}
\vspace{-0.35in}
\caption{An innovative component of the 2DDNA system is the use of automatic discoloration detection in images caused by synthesis and missing oligo errors. A specialized encoding scheme for image data separates the R, G, and B content of the images and places them on different oligos (top row). This allows for using the three channels as a form of replication coding. Smoothness irregularities in one but not in the other two channels are indicative of oligo errors in the former (second row). The resulting discoloration is treated as missing pixels, whitened-out and then ``imputed'' using deep neural network inpainting methods. Further subjective image quality improvements are obtained via enhancement and smoothing (third row). Note that this approach is tailor-made for image data and it mitigates the use of costly error-correcting redundancy. For images with fine facial details, unequal error-protection low-density parity-codes can be used in addition to machine learning approaches, with a redundancy of $<7\%$ for the facial data alone.} \label{fig:Pixie}
\vspace{-0.2in}
\end{figure}

Nick-based recording also enables the creation of 2D storage systems, as the nicks do not have to be necessarily superimposed on native DNA. The 2DDNA model of~\cite{pan2022rewritable} superimposed nick-encoded data on synthetic DNA strands.  Such an approach caters to the need for high-volume storage by encoding information in the DNA content and low-volume rewritable data storage by encoding it in the topological domain. Since the most prevalent data format is image data, the method was specialized to encode images into DNA content and image metadata (ownership information, date of access, steganographic messages) in the form of nicks. Two novel features of this 2DDNA system are a) the use of machine learning methods~\cite{pan2020image} to reconstruct the image in the presence of synthesis and missing oligo errors via a combination of automatic discoloration detection, image inpainting and smoothing (see Figure~\ref{fig:Pixie}), b) the superimposition of rewritable metadata on DNA strands via enzymatic nicking.

We conclude this review of different directions in DNA-based and molecular data storage by describing emerging approaches that aim to increase the size of the DNA alphabet through the use of chemical modifications~\cite{tabatabaei2022expanding} and employ synthetic polymers instead of synthetic DNA~\cite{lutz2015coding,laure2016coding}.

In the former work, the DNA alphabet -- $A$, $T$, $G$, $C$ -- is augmented by chemically modified native bases. A chemical modification is a small group of atoms added to a base so that it does not change its Watson-Crick binding affinity (or, at worst, does not significantly compromise it). The idea is to create ``variants'' of the symbol, say $A_1$, $A_2$, $T_1$, $T_2$, $T_3$, etc, that expand the alphabet size but remain distinguishable when sequenced. The main challenge of this approach is to adapt existing sequencing technologies -- Illumina, ONT or PacBio -- to efficiently discriminate all native and chemically modified symbols. The most promising approaches include learning to classify the bases using raw ion current signals from nanopores and kinetic information from PacBio devices. Another potential drawback of using chemically modified DNA is that PCR random access methods cannot preserve the information encoded in modification unless both strands contain ``matching'' chemical modifications. This problem can be remedied through the use of grids of self-rolled nanotubes that use the negative charge of the DNA sugar-phosphate backbone to control its movement via electronic circuits~\cite{khandelwal2022self}.

In the latter line of work, collections of synthetic polymers (usually two polymers, each assigned to one of two possible bit values) of predetermined and largely different masses are interconnected to form bytes. Chemical bonds are introduced between the bytes to form one information-bearing string which is amenable for separate reading of each byte. Synthetic polymers offer the advantage of lower synthesis costs, although the synthesis process remains sequential. However, there are drawbacks, such as the absence of a PCR-type amplification process and a limited range of natural enzymes capable of working with the polymers. Initially, data retrieval from polymers relied on tandem mass spectrometry~\cite{lutz2015coding}, but recent advancements have focused on the development and utilization of specialized solid-state nanopores~\cite{cao2021decoding}. 

\section{Coding-Theoretic Questions} \label{sec:coding}

As pointed out throughout the text, all components of different DNA storage systems introduce errors. For example, synthesis errors mostly manifest themselves in the form of substitution errors, while errors introduced during nanopore sequencing are standardly modeled as combinations of substitution and indel errors. In addition to these well-studied error models, many previously unexplored research directions in coding theory were purely motivated by molecular storage. Some of these questions and solutions for the same were described in the review paper~\cite{yazdi2015dna1}. To avoid overlaps with the topics in~\cite{yazdi2015dna1}, we hence focus on a sampling of more recent analytic questions pertaining to modeling the DNA storage channel, decoding information via trace reconstruction and designing codes for DNA Punchcard systems.\\ 
 
\textbf{1. A DNA storage channel model}\\ 

Our analysis pertains to the first model of a DNA storage channel, described in~\cite{kiah2016codes}. It provides a simplified, yet conceptually accurate, abstraction of microarray-based synthesis and shotgun-type sequencing. To facilitate the mathematical exposition, we start with some relevant terminology.

Let $n$ be a positive integer, $[q]=\{{0,1,\ldots,q-1\}}$, and $\textbf{x} \in [q]^n.$ Choose a constant integer $0<\ell$. The \emph{$\ell$-profile vector} of $\textbf{x}$, $\pi_{\ell}(\textbf{x})$, is a vector of length $q^{\ell}$ whose coordinates are indexed by all possible $q$-ary strings of length $\ell$, in lexicographic order. The $i$-th entry of $\pi_{\ell}(\textbf{x})$ is the number of substrings of 
$\textbf{x}$ equal to the $i$-th string in the lexicographical order of strings in $[q]^{\ell}$. Note that the sum of entries of a profile vector of a string of length $n$ equals $n-\ell+1$. 

For example, if $q=2$, $n=5$, $\ell=2$, and $\textbf{x}=11011$, then 
$\pi_2(\textbf{x})=0112$, and $0+1+1+2=4=5-2+1$. The lexicographical ordering used is $(00,01,10,11)$.

For simplicity of notation, we henceforth drop the subscript $\ell$ as it will be made clear from the context.

We say that $\pi()$ is a \emph{valid profile} if there exists a string with that given profile. Otherwise, we say that the profile is not valid. For the example parameters above, $\pi=2002$ is not a valid profile, since there is no binary vector of length $n=5$ that contains $2$ substrings $00$ and $2$ substrings $11$.

Observe that two different strings can share the same profile. For example, consider the following collection of strings
\[\{{0000,0010,0100,0110,1001,1011,1101,1111\}}.\]
The $\ell=2$-mer equivalence classes of the strings, with two strings being equivalent if their $\ell=2$-mer profile vectors are the same, are
\[\{0000\}, \{0010,0100,1001\},\{0110,1011,1101\},\{1111\}.\]
Clearly, the profile vectors of the four equivalence classes are $(3,0,0,0)$, $(1,1,1,0)$, $(0,1,1,1)$ and $(0,0,0,3)$.

Next, for profile vectors of two $q$-ary strings $\textbf{x}$ and $\textbf{y}$, let us define their
\emph{asymmetric profile distance} according to 
$$\Delta(\textbf{x},\textbf{y})=\max\{{\partial(\textbf{x},\textbf{y}),\partial(\textbf{y},\textbf{x})\}},$$
where $\partial(\textbf{x},\textbf{y})=\sum_{i=0}^{q^{\ell}-1}\,\max\{{\pi(\textbf{x})_i-\pi(\textbf{y})_i,0\}}$ and the subscript $i$ denotes the $i$-th coordinate of the vector.

The question of interest is to design the largest possible codebook of $q$-ary vectors, $\C_{q,n,\ell,d},$ of length $n$ such that the minimum pairwise asymmetric distance of their $\ell$-profile vectors is at least $d$. For $d \geq 1$, one has to automatically preclude simultaneously including two strings from the same equivalence class in $\C_{q,n,\ell,d}$, since in that case, the profiles of the strings are the same.  

\begin{figure}%{l}{0.5\textwidth}
% \vspace{-0.1in}
\begin{center}
\includegraphics[width=0.8\textwidth]{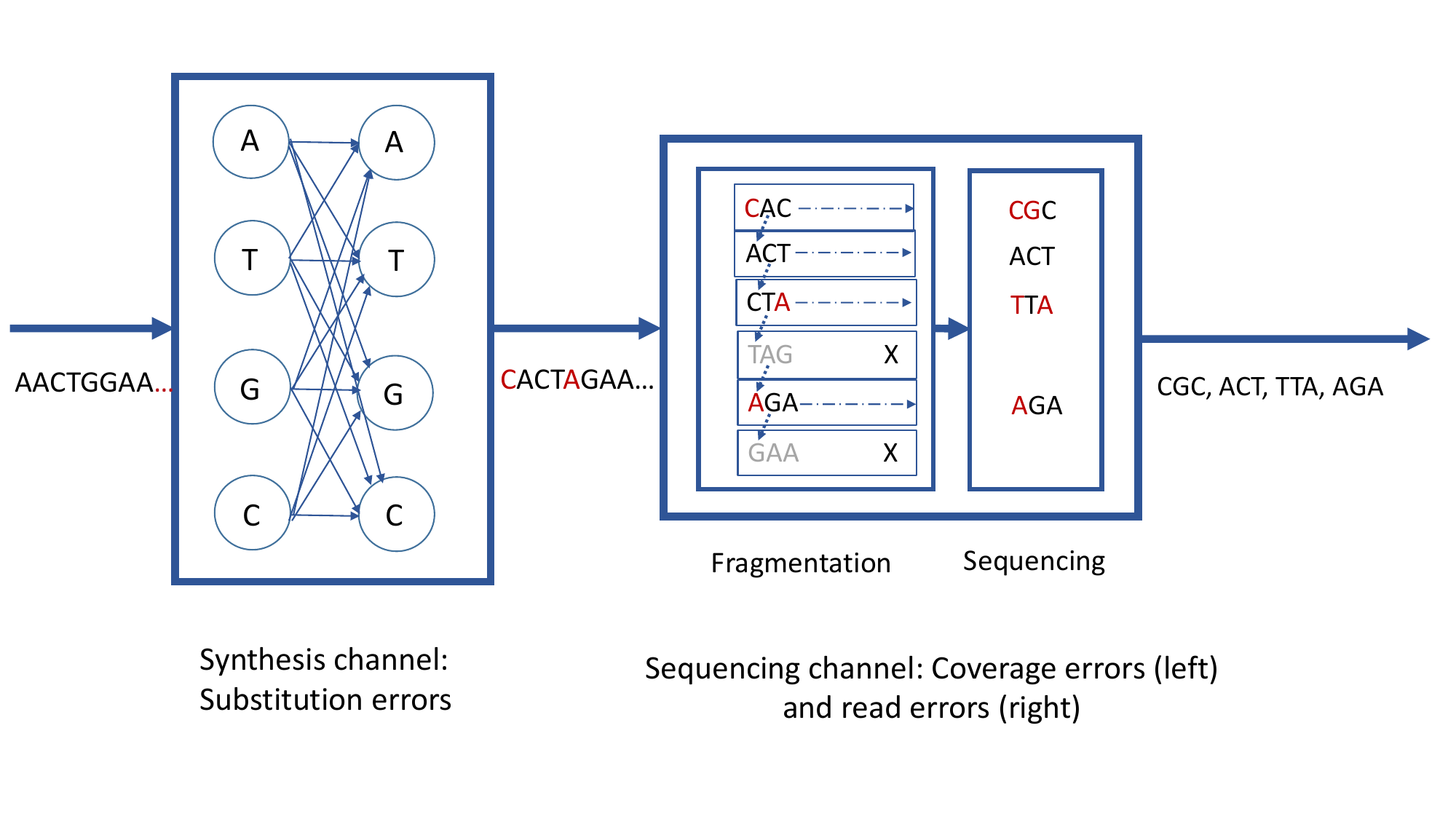}
\end{center}
\vspace{-0.4in}
\caption{A DNA-based data storage channel model. The input into the channel is a string of length $n$ over the DNA alphabet $\{{A,T,G,C\}}$. The output of the channel comprises a collection of \emph{noisy substrings} of the input string. Noise is introduced at three different stages of the write-read process. During synthesis (leftmost panel), one encounters \emph{synthesis substitution errors}. In the figure, such errors are marked in red, and they include the bases $C$ and $A$. Note that synthesis errors propagate through the channel as they are ``imprinted'' into the string that is to be sequenced. Once the string is synthesized, it is read by first fragmenting it into $\ell$-mers (in the example, $3$-mers), some of which may be missing due to \emph{coverage errors} (marked in gray). The substrings are then read through synthesis, and the reading process itself can lead to the introduction of additional \emph{sequencing substitution errors} within the substrings. The output of the channel is an incomplete, noisy collection of substrings of the input string.} \label{fig:dnachannel}
\vspace{-0.1in}
\end{figure}

The motivation for studying the previously introduced problems comes directly from the models of DNA-based data storage depicted in Figure~\ref{fig:dnachannel}. When sequencing DNA, the input to the sequencer is a string, while the output is a collection of substrings (reads) generated (for this model) by Illumina sequencing technologies. The asymmetric distance allows one to account for three types of errors, \emph{synthesis substitution errors,} whose number is assumed to be upper-bounded by $s$; \emph{coverage errors,} modeled as missing substrings, the number of which does not exceed $c$; \emph{sequencing substitution errors,} which manifest themselves as individual substitution errors within the substrings (reads), and the number of which does not exceed $o$. If the minimum asymmetric distance of the profiles of the codestrings satisfy $d_{min} \geq 2s+c+o$, then the code can correct the corresponding number of synthesis, coverage and sequencing errors. Several abstractions are made to make the analysis of this model tractable. First, it is assumed that one can perfectly count distinct oligo types. In the example with $\textbf{x}=11011$ and $\pi(\textbf{x})=0112$, it is assumed that one can determine that there were \emph{two distinct} substrings $11$ in the original string. In practice, Illumina systems actually do report all sequenced oligos, but without the information if these oligos are replicas of the same substring or replicas of multiple identical substrings. This issue can be mitigated through the use of long-read technologies which are known to resolve issues with repeats. Second, synthesis errors are usually context-dependent, and repeats make the process difficult or outright impossible. To make the model more realistic, we would require the codestrings to be repeat-free, but this would make the subsequent analysis very hard. Third, since multiple replicas of the same string may be generated during sample preparation, and each of these strings can be subject to different error-patterns, one substring could give rise to multiple erroneous substrings. How many replicas are present depends on the coverage depth. and the replica count being above one is expected to be a small probability event.

Given that the DNA storage channel accepts \emph{strings} at the input and produces profiles at the output, it is not immediately clear how to ensure that minimum asymmetric distance constraints are met in the substring profile domain while working directly with the input strings. The key ideas for solving this problem are described in the exposition to follow. 

The connection between strings and their profiles is established through the use of \textbf{de Bruijn graphs}~\cite{bruijn1975acknowledgement,Bollobas:1998}.

A \emph{directed graph (digraph)} $D$ is a pair of sets $(V,E)$, where $V$ is the set of \emph{nodes}
and $E$ is a set of ordered pairs of $V$, termed \emph{arcs}.
If $e=(v,v')$ is an arc, we call $v$ the {\em initial} node (tail) and $v'$ the {\em terminal} (head) node.
We allow loops (i.e., we allow $v=v'$) and multiple arcs between nodes.

The \emph{incidence matrix} of a digraph $\textbf{D}$ is a matrix $\textbf{B}(\textbf{D})$ in $\{-1,0,1\}^{V\times E}$, where
{\small
\begin{equation*}
\textbf{B}(\textbf{D})_{v,e}=
\begin{cases}
\,\;\;1, & \mbox{if $e$ is not a loop and $v$ is its terminal (head) node},\\
-1, & \mbox{if $e$ is not a loop and $v$ is its initial (tail) node},\\
\,\;\;0, & \mbox{otherwise}.
\end{cases}
\end{equation*}
}

\begin{figure}%{l}{0.5\textwidth}
% \vspace{-0.1in}
\begin{center}
\includegraphics[width=0.7\textwidth]{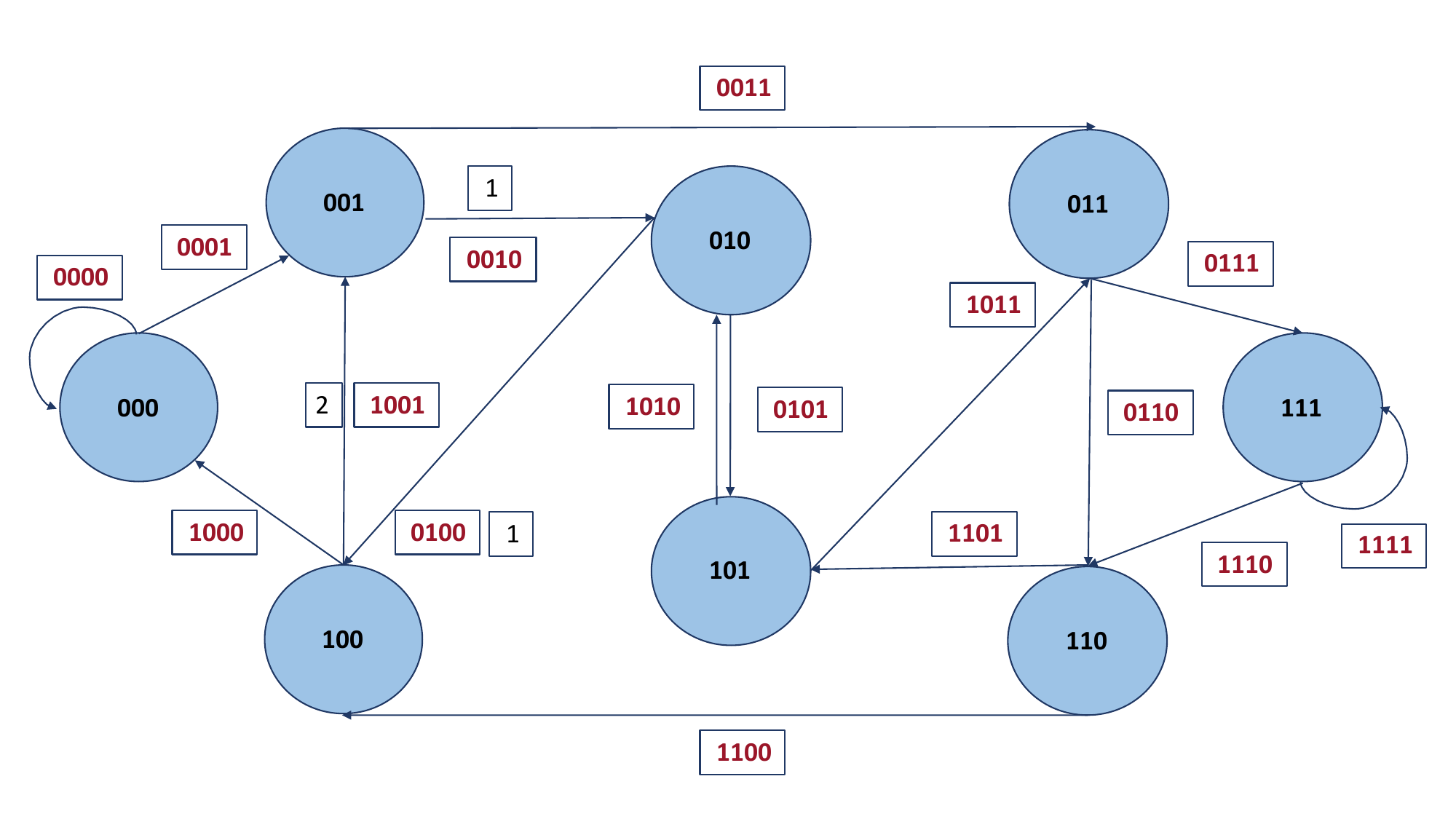}
\end{center}
\vspace{-0.1in}
\caption{The de Bruijn graph for $q=2, \, \ell=4$. Vertex labels are binary vectors of length $\ell-1$, while arc labels are binary vectors of length $\ell$. The nonnegative integer weight of arcs describe the $\ell=4$-mer profile vector of the string $\textbf{x}=1001001$, $\pi(\textbf{x})=00101000002000000$. Unweighted arcs are assumed to have weight $0$.} \label{fig:db}
\vspace{-0.2in}
\end{figure}

Given $q$ and $\ell$, the (standard) {\em de Bruijn graph} is defined on the node set $[q]^{\ell-1}$, where once again we use $[q]=\{{1,\ldots,q\}}$. For $\vv,\vv'\in\bbracket{q}^{\ell-1}$, the ordered pair $(\vv,\vv')$ belongs to the arc set if and only if 
$v_i=v'_{i-1},$ for $2\le i\le\ell-1$. We label the arc $(\vv, \vv')$ with the length-$\ell$ string $\vv v'_{\ell-1}$, and when needed, refer to the arcs by invoking their labels. As an example, let $q=2, \, \ell=4$ and consider the de Bruijn graph in Figure~\ref{fig:db}. The nodes $\vv=101$ and $\vv'=010$ are connected by the arc $1010$ which originates from $\vv$
and terminates in $\vv'$ as the suffix of $\vv$ of length $\ell-2=2$ equals $01$, which is also the prefix of length $\ell-2$ of $\vv'$.  

The notion of de Bruijn graphs can be extended to preclude the presence of certain $\ell$-mer arc labels or $(\ell-1)$-mer vertex labels~\cite{ruskey2012bruijn}. For such \emph{restricted de Bruijn graphs,} the set of 
allowed $(\ell-1)$-mers is denoted by $S$. The corresponding restricted de Bruijn graph is denoted by $D(S)$. The importance of restricted de Bruijn graphs for DNA-based storage systems lies in the fact that $S$ may be chosen to satisfy additional sequence constraints, such as balanced $GC$ constraint (i.e., balanced $\ell-1$-mers and balanced $S$-sets). For $q=2$, ``balanced'' refers to the corresponding substrings containing the same number of $0$s and $1$s, while for larger values of $q$, it is used to denote balanced (or nearly balanced) $GC$ content.

A \emph{walk} of length $n$ in a digraph is an ordered sequence of nodes
$v_0v_1\cdots v_n$ such that $(v_i,v_{i+1})\in E$ for all
$i\in\bbracket{n}$. A walk is \emph{closed} if $v_0=v_n$. A \emph{cycle} is a closed walk with distinct nodes, i.e., $v_i\ne v_j$, for
$0\le i<j< n$. A loop is a cycle of length one.  
Given a subset $A$ of the arc set, let $a\in \{0,1\}^{|E|}$ be its \emph{incidence vector}, so that $a_e=1$ if $e\in A$ and $a_e=0$ 
otherwise. For the incidence vector $a$ of a closed walk in $D$, we have
$\bf{B}(D)a=\vzero$. A closed walk is {\em Eulerian} if it includes all arcs in $E$. %A cycle is {\em Hamiltonian} if it includes all nodes in $V$.
A digraph is \emph{strongly connected} if for
all $v, v' \in V$, there exists a walk from $v$ to $v'$ and vice versa. Furthermore, a necessary and sufficient condition for a strongly connected graph to have a closed Eulerian walk is that, for each node, the number of incoming arcs is equal to the number of outgoing arcs. 

The de Bruijn digraphs of interest to our problem have arcs weighted by nonnegative integers. The weight of an arc indicates the count of its substring label in a string. As an example, in Figure~\ref{fig:db}, only three arcs have integer labels marked in black. All other arcs are assumed to have weight zero. Since the label of each arc is uniquely determined by the source and terminal vertex, one can omit the sequence label and only retain the arc weights. The weights in the figure describe how many times an arc has to be traversed and simultaneously capture the profile of a string. In this setting, one of the possible strings whose profile is shown in the figure equals $1001001$, since it contains the following substrings of length $\ell=4$: $\{{1001,0010,0100,1001\}}$. Hence, to recover the string, the arc labeled $1001$ has to be traversed twice, while the arcs labeled $\{{0010,0100\}}$ have to be traversed once. The length of the string is $n=4+\ell+1=7,$ since there are $4$ substrings in total. The length of the walk in the graph equals the sum of the arc labels or the number of substrings of the string, which equals $n-\ell+1=4$.

A walk from some node $\vv$ to another node $\vv'$ in $\textbf{D}(S)$ describes a string that starts with $\vv$ and ends with $\vv'$ and whose $\ell$-mers belong to $S$. {\em Closed strings} are strings that start and end with the same $(\ell-1)$-mer and they correspond with closed walks in $\textbf{D}(S)$. Strings corresponding to walks of length $n-\ell+1$ in $\textbf{D}(S)$ (and, consequently, profiles of strings of length $n$) are denoted by $\Q(n;S)$. At the same time, $\Qb(n;S) \subseteq \Q(n;S)$ is used to denote the subset of \emph{closed strings}. The set of profile vectors of closed strings is denoted by $\pQb(n;S)$. The reason for introducing closed strings is that a number of counting arguments simplify in this case, while the resulting code rate for constant $\ell$ remains largely unaffected.

Suppose that $\vu \in \pQb(n;S)$. Then, the {\em flow conservation equations} below have to hold:
\begin{equation}\label{eq:flow}
\vB(\textbf{D}(S))\vu=\vzero.
\end{equation}
Furthermore, let $\vone$ denote the all-ones vector. Since the number of $\ell$-mers in a string of length $n$ equals $n-\ell+1$, 
we also have
\begin{equation}\label{eq:sum}
 %\sum_{\vz\in S} u_\vz=n-\ell+1.
 \vone^T\vu=n-\ell+1,
\end{equation}
where $T$ denotes the transpose operator. Let $\vA(S)$ be $\vB(\textbf{D}(S))$ augmented with a top row $\vone^T$; also, let $\vb$ be a vector of length $|V(S)|+1$ with a one as its first entry, and zeros elsewhere.
Equations \eqref{eq:flow} and \eqref{eq:sum} may then be rewritten as
$$\vA(S)\vu=(n-\ell+1)\vb.$$

Consider the following two definitions of sets of integer points:
\begin{align}
\mathcal{F}(n;S) & \triangleq\{\vu\in \ZZ^{|S|}: \vA(S)\vu=(n-\ell+1)\vb,\ \vu\ge\vzero\}, \label{sol_polytope}\\
\mathcal{E}(n;S) &\triangleq\{\vu\in \ZZ^{|S|}: \vA(S)\vu=(n-\ell+1)\vb,\ \vu>\vzero \} \label{sol_interior}.
\end{align}
It is straightforward to see that the profile vector of any
closed string must belong to $\mathcal{F}(n;S)$.  Conversely, any vector in $\mathcal{E}(n;S)$ is a profile 
vector of some string in $\Qb(n;S)$. 

The above formulation can be used to establish a count of the number of profile equivalence classes as follows. Suppose $D(S)$ is strongly connected. Then, under certain mild constraints and for a constant $\ell$, it can be shown that $|\E(n;S)| \sim n^{|S|-|V(S)|}$ and $|\F(n;S)| \sim n^{|S|-|V(S)|}$, and that as a result $|\pQb(n;S)| \sim n^{|S|-|V(S)|}$. Here, the symbol $\sim$ is used to indicate that for sufficiently large $n$, the sizes of the sets scale as the term on the right\footnote{A rigorous statement of the results involves the definition of quasipolynomials and is therefore omitted.}. In a nutshell, the result follows by counting the solutions of the defining conditions for points in $\mathcal{F}(n;S)$ and $\mathcal{E}(n;S)$ via \emph{lattice point enumeration techniques} and what is known as Erhart-McDonald's reciprocity theory. The Erhart-McDonald's reciprocity theory is a broad generalization of a more simple result known as \emph{Pick's theorem,} which expresses the area of a polygon in terms of the number of lattice points in its interior~\cite{pick1899geometrisches}.

For unrestricted de Bruijn graphs, which are strongly connected, we have $n^{|S|-|V(S)|}=n^{q^{\ell}-q^{\ell-1}}.$
This expression is the asymptotic for the number of \emph{distinct} $\ell$-gram profiles of $q$-ary strings of length $n$. 
It also establishes that $|\C_{q,n,\ell,d=1}| \sim n^{q^{\ell}-q^{\ell-1}}$.

To determine $|\C_{q,n,\ell,d}|$ for $d>1$, we need to ensure that the profiles are not only distinct but at an asymmetric distance $d$ from each other. This can be ensured by adding more constraints to the profile vectors (i.e., in addition to the flow- and sum-constraints) that also take the form of linear equations. The solution involves using \emph{Varshamov codes}~\cite{varshamov1973class}, designed specifically for asymmetric channels. For convenience, we describe these codes below.

Fix $d$ and let $p$ be a prime such that $p>\max\{{d,N\}},$ where for notational convenience we use $N$ to stand for $|S|$. Next, choose $N$ distinct nonzero elements $\alpha_1,\alpha_2,\ldots,\alpha_N$ in $\ZZ/p\ZZ$ and consider the
matrix 
\[
\vH\triangleq
\left(\begin{array}{cccc}
\alpha_1 & \alpha_2 & \cdots & \alpha_N \\
\alpha_1^2 & \alpha_2^2 & \cdots & \alpha_N^2 \\
\vdots &\vdots& \ddots &\vdots\\
\alpha_1^d & \alpha_2^d & \cdots & \alpha_N^d 
\end{array}\right).
\]
Pick any vector $\vbeta\in(\ZZ/p\ZZ)^d$ and define a code
%\textcolor{orange}
{
\begin{equation} \label{eq:varshamov}
\C(\vH,\vbeta)\triangleq\{\vu:\vH\vu\equiv\vbeta \bmod p\}.
\end{equation}}
Then, $\C(\vH,\vbeta)$ is an asymmetric error-correcting codes of length $N$ and (designed) minimum asymmetric distance $d+1$. Hence, all the codestrings of a Varshamov code that are valid profile vectors are also 
$d$-asymmetric-error-correcting codestrings. More precisely, we have $|\C_{q,n,\ell,d}| \sim |\C(\vH,\vbeta)\cap \pQ(n;S)|,$ for all $\vbeta\in(\ZZ/p\ZZ)^d$. By the pigeon-hole principle, there exists a $\vbeta$ such that $|\C(\vH,\vbeta)\cap \pQ(n;S)|$ is at least $|\pQ(n;S)|/p^d$. The choice of $\vbeta$ that guarantees this lower bound is not known but this does not impose consequential practical issues. Furthermore, $|\C\cap\pQ(n;S)|$ is typically strictly smaller than $|\C|$, and deriving analytical bounds for the code size is nontrivial (see~\cite{kiah2016codes} for details).

Suppose that $\C$ is a Varshamov asymmetric distance error-correcting code with parameters $(N,d)$. We construct DNA profile codes from $\C$ as follows:

\begin{enumerate}
\item When $N=|S|$, we intersect $\C$ with $\pQ(n;S)$ to 
obtain an $\ell$-gram asymmetric error-correcting code. Simply put,
we pick out the codestrings in $\C$ that are also profile vectors.

\item When $N<|S|$, we extend each codestrings in $\C$ to 
a profile vector of length $|S|$ in $\pQ(n;S)$. 
In contrast to the previous construction, in this case we may in principle obtain a code with the same cardinality as $\C$. However, one may not always be able to extend an arbitrary string to a profile vector. 
\end{enumerate}

As an example, let $q=2$, $\ell=3$, $S=\{001,010,011,100,101,110\}$ so that $N=6$. Let $d=3$ and choose $p=7$, so that
\[
\vH=
\left(\begin{array}{ccc ccc}
1 & 2 & 3 & 4 & 5 & 6 \\
1 & 4 & 2 & 2 & 4 & 1
\end{array}\right)\mbox{ and }
\vbeta=\left(\begin{array}{c}
0\\0
\end{array}\right).
\]
Then, $\C(\vH,\vbeta)$ contains the following strings.
\begin{center}
\begin{tabular}{cccccc}
$(4, 0, 0, 1, 0, 1)$ 
&&&$(0, 1, 1, 4, 0, 0)$\\

${\color{blue} (2, 2, 0, 2, 0, 0)}$ & $\leftrightarrow$ & {\color{blue}00100100}  
&$(0, 1, 0, 0, 4, 1)$\\

$(1, 4, 0, 0, 1, 0)$
&&&$(0, 0, 4, 1, 1, 0)$\\

${\color{blue} (1, 1, 1, 1, 1, 1)}$ & $\leftrightarrow$ & {\color{blue}00101100}  
&${\color{blue} (0, 0, 2, 0, 2, 2)}$ & $\leftrightarrow$ & {\color{blue}01101101} \\ 

$(1, 0, 1, 0, 0, 4)$\\
\end{tabular}
\end{center}
Of these Varshamov codestrings, only three (marked in blue) are valid profile vectors from $\pQ(8;S)$. Hence, for this setting of parameter values, we would only have a codebook of three allowed profile codestrings. 

Two final observations are in place. First, counting the codestrings in a Varshamov-type profile code once again, like in the equivalence class counting framework, reduces to computing the lattice point enumerator of the intersection of the lattices defined by $\vA(S)$ and $\C(\vH,\vbeta)$. Finding lattice point enumerators is a fundamental problem in discrete optimization and high-quality software suites for solving the problem are available. One such software, {\tt LattE}, introduced in~\cite{baldoni2014user}, is based on an elegant algorithm described in~\cite{barvinok1999algorithmic} that triangulates the supporting cones of the vertices of a polytope to obtain simplicial cones that are then recursively decomposed into unimodular cones. The algorithm performs enumeration of lattice points in polynomial time
whenever the dimension of the polytope is fixed. Second, the asymmetric code construction procedure was implemented to produce profile vectors, which are actual outputs of the channel and not the desired input codestrings. We hence need to convert the profiles back into strings, with exactly one string corresponding to one profile. This can be accomplished by using de Bruijn graphs that capture both the substrings and their multiplicities -- all that is needed is to find a path in the graph that traverses each arc a number of times indicated by its weight multiplicity. This is akin to the process of sequence assembly that is widely used in computational biology~\cite{nagarajan2013sequence}.
%%%%%%%%%%%%%%%%%%%%%%%%%%%%%%%%%%%%%%%%%%%%%%%%%%%%
%Codes for unique reconstruction (no repeats);
%%%%%%%%%%%%%%%%%%%%%%%%%%%%%%%%%%%%%%%%%%%%%%%%%
%%%%%%%%%%%%%%%%%%%%%%%%%%%%%%
%%%%%%%%%%%%%%%%%%%%%%%%%%%%%%%%%%%%%%%%%%%%%%%%%
\clearpage
\textbf{2. Coded Trace Reconstruction}\\

The problem of trace reconstruction was introduced in~\cite{batu2004reconstructing}, motivated by sequence analysis problems by Levenshtein~\cite{levenshtein2001efficient} and practical sequence alignment questions in phylogeny and computational biology. The relevance of the problem to DNA-based data storage comes from the connection to sequence alignment, which is necessary when reading the information content via nanopore sequencers (see Section~\ref{sec:platforms}). There, we discussed various sequence alignment algorithms from computational biology that can be used to create consensus sequences from multiple noisy reads. Unfortunately, most of these methods rely on dynamic programming approaches, which are notoriously hard to analyze. Trace reconstruction, on the other hand, is an abstraction that is conceptually simple to state and understand, but also coupled with nontrivial analytical challenges. 

In a nutshell, the trace reconstruction problem asks how many noisy reads are needed to reconstruct the original string through the use of a simple-to-analyze consensus algorithm. More formally, the assumption is that there exists an unknown string $\textbf{x}\in\{0,1\}^n$, and that one is given access to \emph{traces} of $\textbf{x}$, which are generated by passing $\textbf{x}$ through a deletion channel (i.e., a nanopore). The deletion channel independently deletes bits of $\textbf{x}$ with a given deletion probability $q$, and each pass through the channel produces a trace that is independent of all other traces. Clearly, traces represent nothing more than subsequences of $\textbf{x}$ of varied length formed in a probabilistic, i.i.d manner. The formal goal is to minimize the number of traces, i.e., the number of reads, that need to be acquired in order to reconstruct $\textbf{x}$ with high probability. 

Note that the binary trace reconstruction setting is significantly more difficult than the corresponding larger-alphabet problem. This is why we mostly focus on results for binary strings. As will be explained towards the end of the section, results for binary strings can be easily translated into results for quaternary strings. 

The focus of the trace reconstruction research area has been mostly on two types of approaches: \emph{worst-case}~\cite{de2017optimal,nazarov2017trace}, where the requirement is for the reconstruction procedure to work for all strings in $\{0,1\}^n$, and \emph{average-case}~\cite{holden2018subpolynomial,hartung2018trace,peres2017average}, where the reconstruction algorithm is only required to work with high probability for a string selected uniformly at random. Formally, worst-case trace reconstruction is concerned with designing a reconstruction algorithm, $\mathcal{R},$ such that \emph{for every} $\textbf{x}\in \{{0,1\}}^n$ one has
\begin{equation*}
P_{T_1,\dots,T_t}\left[\mathcal{R}(T_1,\dots,T_t)=\textbf{x}\right]\geq 1-1/n,
\end{equation*}
where $T_{i}, \, i \in [t],$ stands for traces of $\textbf{x}$ which are i.i.d. according to the output distribution of a deletion channel with deletion probability $q$. The goal is to make $t=t(n)$ as small as possible. For average-case reconstruction, we require that $\mathcal{R}$ satisfy 
\begin{equation}\label{eq:avgtr}
2^{-n}\sum_{\textbf{x}\in \{{0,1\}}^n} P_{T_1,\dots, T_t}\left[\mathcal{R}(T_1,\dots,T_t)=\textbf{x}\right]\geq 1-1/n,
\end{equation}
where we would once again like to make $t=t(n)$ as small as possible, and where the traces $T_{i}, \, i \in [t],$ have the same properties as stated for the worst-case problem. It is clear that the number of traces required for average-case trace reconstruction is smaller than that required for worst-case trace reconstruction. The state-of-the-art results for average-case reconstruction~\cite{holden2018subpolynomial} established that $\exp(O(\log^{1/3} n))$ traces suffice to reconstruct a random $n$-bit string under arbitrary \emph{constant} deletion probability $q$. Handling the worst-case setting is much more challenging. The current best upper bound~\cite{chase2020new} equals $\exp(O(n^{1/5}))$, improving the results of~\cite{de2017optimal,nazarov2017trace} $\exp(O(n^{1/3}))$ based on an algorithm that exploits \emph{single-bit statistics} of the traces\footnote{The latter upper bound is tight for single-bit statistics algorithms.}. The gap between the upper and lower bound is still prohibitively large. The state-of-the-art lower bounds~\cite{chase2021new} are roughly $\log^{5/2}\,n\log\log^{-7}n$ traces for average-case trace reconstruction and $n^{3/2}\log^{-7}n$ traces for the worst-case setting. For some more recent results regarding $k$-mer statistics approaches, please see~\cite{mazooji2023substring,cheng2023k}.

Unlike other applications, DNA-based data storage allows choosing a subset of strings with desirable properties to be used for trace reconstruction. For example, one can focus on fixed-weight strings, strings that satisfy bounded runlength (homopolymer) properties, balanced $GC$ contents and others. This more restricted sequence selection process naturally leads to the question of \emph{coded trace reconstruction}~\cite{cheraghchi2020coded}. Here, the goal is to design codes with asymptotic rate equal to $1$ that are also efficiently encodable and decodable using significantly fewer traces than needed for the unrestricted setting. As in all other works, the assumption is that we work with constant channel deletion probabilities. A simple yet significantly more parameter-restricted line of work addressed the coded trace reconstruction problem for a \emph{constant number} of deletions, using concatenations of Varshamov-Tenengolts codes~\cite{abroshan2019coding}. Another line of work~\cite{brakensiek2020coded} built upon the techniques of~\cite{cheraghchi2020coded} and provided improvements on the number of traces required as a function of the rate. A drawback of the latter method is the requirement of preprocessing which requires superpolynomial time.

To better understand our approach to coded trace reconstruction, let us revisit the ideas from~\cite{yazdi2017portable} which for the first time modeled the nanopore sequencing process as trace reconstruction. The codestrings were designed to satisfy block-wise $GC$-balancing constraints, with each block of $8$ symbols in $\{{A,T,G,C\}}$ perfectly balanced. Balancing constraints were used to ensure correct synthesis, but somewhat serendipitously proved useful for trace reconstruction. The utility of balancing for string reconstruction is in part due to the related runlength constraints. More precisely, block-based balancing enforces symbol 
runlength (homopolymer) constraints: for an arbitrary collection of balanced blocks of size $b$, the runlength of each symbol is upper bounded by $b$. The traces obtained via nanopore sequencing in~\cite{yazdi2017portable} were used to form a consensus sequence, which was then updated in several iterations by checking if the block-level balancing constraints are met. For simplicity, we will illustrate the underlying ideas through a simple adaptation of the Bitwise Majority Alignment (BMA) algorithm~\cite{batu2004reconstructing}, although this algorithm does not perform as well as the actual algorithm used in~\cite{yazdi2017portable} due to not being able to handle context-dependent indels and substitution errors. 

As an example, let the codestring to be sequenced by nanopores equal 
$$\textbf{s}=AATGGCGA \; \;  TTCCGGAA \; \; GGGAATCA,$$
comprising three blocks of length $8$, each with a perfectly balanced ($50\%$) $GC$ content (note that the string is parsed into blocks for ease of visualization). Now, assume that the sequencer produced $5$ reads/traces based for the input string $\textbf{s}$, as listed below.

\begin{equation}
\begin{aligned}
&ATGGCGTTCGGAAGGATCA\\
&AATGGTTCCGGAAGGAAT\\
&AATGGCGATTCCGGGGGAAA\\
&GCGATTCCGGGGAATA\\
&ATGGATCGAGGATCA.
\end{aligned} \notag
\end{equation}

The algorithm proceeds by constructing the consensus sequence by focusing on one position at the time, finding the majority symbol and calling it the consensus symbol, and then shifting the mismatched symbols one position to the right. The first three steps of the approach applied to the above DNA strings are presented below, along with the final consensus result. Ties are broken arbitrarily but recorded for subsequent reexamination. Majority symbols are written in blue, while minority symbols are replaced by ``-'' and moved to the right. Note that in Step 2, the tie is broken in favor of $A$, and both symbols $A$ and $T$ are recorded for later use.
\clearpage
Step 1:\\
\begin{equation}
\begin{aligned}
&ATGGCGTTCGGAAGGATCA\\
&AATGGTTCCGGAAGGAAT\\
&AATGGCGATTCCGGGGGAAA\\
&\textcolor{blue}{-}GCGATTCCGGGGAATA\\
&ATGGATCGAGGATCA\\
&\textcolor{blue}{A.}
\end{aligned} \notag
\end{equation}

Step 2:\\
\begin{equation}
\begin{aligned}
&A\textcolor{blue}{-}TGGCGTTCGGAAGGATCA\\
&AATGGTTCCGGAAGGAAT\\
&AATGGCGATTCCGGGGGAAA\\
&\textcolor{blue}{-}\textcolor{blue}{-}GCGATTCCGGGGAATA\\
&A\textcolor{blue}{-}TGGATCGAGGATCA\\
&\textcolor{blue}{A}\textcolor{blue}{A} \\
&\textcolor{blue}{A}\textcolor{blue}{(A/T).} 
\end{aligned} \notag
\end{equation}

Step 3:\\
\begin{equation}
\begin{aligned}
&A\textcolor{blue}{-}TGGCGTTCGGAAGGATCA\\
&AATGGTTCCGGAAGGAAT\\
&AATGGCGATTCCGGGGGAAA\\
&\textcolor{blue}{-}\textcolor{blue}{-}\textcolor{blue}{-}GCGATTCCGGGGAATA\\
&A\textcolor{blue}{-}TGGATCGAGGATCA\\
&\textcolor{blue}{A}\textcolor{blue}{A}\textcolor{blue}{T} \\
&\textcolor{blue}{A}\textcolor{blue}{(A/T)\textcolor{blue}{T}.} 
\end{aligned} \notag
\end{equation}
The original sequence, the consensus sequence, and the consensus with ties are summarized below, respectively. Mismatches are indicated in red, and the sequences are parsed into groups of $8$ symbols for future analysis.
\begin{equation}
\begin{aligned}
&AATGGCGA \; \; \;\;\;\;\;\;\, TTCCGGA\textcolor{red}{A} \; \; G\textcolor{red}{G}G\textcolor{red}{A}AT\textcolor{red}{CA} \\
&\textcolor{blue}{A}\textcolor{blue}{A}\textcolor{blue}{T}\textcolor{blue}{GGCGA \; \;\;\;\;\;\;\;\, TTCCGGA\textcolor{red}{G} \; \; G\textcolor{red}{A}G\textcolor{red}{G}AT\textcolor{red}{ACAT}} \\
&\textcolor{blue}{A}\textcolor{blue}{(A/T)}\textcolor{blue}{T}\textcolor{blue}{GGCGA \; \;  TTCCGGAG \; \; (A/G)AGGATACAT.} 
\end{aligned} \notag
\end{equation}
Let us now examine the consensus sequence in the middle row. Clearly, the consensus is longer than the original string, which is a consequence of the right-shift process for minority symbols. Furthermore, the second block is disbalanced, as there is one more $GC$ symbol than allowed, but there were no tie-breaks at that particular location that can help resolve the problem. Furthermore, since the previous block of $8$ symbols was balanced, it is reasonable to assume that the ``boundary'' of the second and third blocks have shifted due to alignment errors. Looking ahead for the first appearance of an $AT$ symbol in the consensus, we can try to ignore all symbols between the last symbol that causes a disbalance and the first occurrence of a symbol of the correct type. Note that since we had a tie for the first symbol in the first block ($A$ versus $G$), it is advisable to change the break of tie to avoid excluding one extra symbol. This leads to the following modification of the consensus string.
\begin{equation}
\begin{aligned}
&AATGGCGA \; \; \;\;\;\;\;\;\, TTCCGGAA \; \; \textcolor{red}{G}GGA\textcolor{red}{AT}CA \\
&\textcolor{blue}{A}\textcolor{blue}{A}\textcolor{blue}{T}\textcolor{blue}{GGCGA \; \;\;\;\;\;\;\;\, TTCCGGAA} \; \; \textcolor{blue}{\textcolor{red}{A}GGA\textcolor{red}{TA}CA\textcolor{red}{T}}. \\
%&\textcolor{blue}{A}\textcolor{blue}{(A/T)}\textcolor{blue}{T}\textcolor{blue}{GGCGA \; \;  TTCCGGAG \; \; (A/G)AGGATACAT.} 
\end{aligned} \notag
\end{equation}
The last block of $9$ symbols in the consensus is obviously erroneous since there is one more $AT$ symbol present than as expected and the block is of length $9$ rather than $8$. 

The example motivates the ideas to be pursued for code constructions that offer provable performance guarantees for trace reconstruction algorithms, which for the best results need to be more sophisticated than simple BMA-type methods. The key insight is to group symbols into blocks with constraints such as balanced content (or runlength constraints) such as above and ensure that the boundaries of the blocks can be determined with high probability. An additional layer of protection may be added to correct errors in the blocks whenever the deletion probability is sufficiently high. The construction, as well as the main results for coded trace reconstruction are formally described next. 

Given a code $\cC\subseteq \{{0,1\}}^n$, we say that $\cC$ can be \emph{efficiently reconstructed from $t(n)$ traces} if there exists a polynomial $p(n)=\Omega(n)$ and a polynomial-time algorithm $\mathcal{R}$ such that for every $\textbf{c}\in\cC$ one has
\begin{equation*}
P_{T_1,\dots,T_t}\left[\mathcal{R}(T_1,\dots,T_t)=c\right]\geq 1-1/p(n),
\end{equation*}
where the traces $T_i, \, i \in [t],$ are i.i.d. according to the output distribution of the deletion channel with deletion probability $q$ on input $\textbf{c}$. This definition corresponds to the worst-case trace reconstruction problem restricted to codestrings of $\cC$.
The goal of coded trace reconstruction is to design \emph{efficiently encodable} codes $\cC$ that can be efficiently reconstructed from $t(n)$ traces for $t(n)$ as small as possible. As a remark, we require a reconstruction success probability $1-1/p(n)$ in order to be able to compare the results of coded reconstruction with those of unrestricted trace reconstruction.

At a high level, the simplest construction splits an $n$-bit message into shorter blocks of length $O(\log^2 n)$, 
encodes each block with an inner code satisfying a certain constraint (such as a runlength/balaning and/or more general error-correcting constraints), and adds markers of length $O(\log n)$ between the blocks. Markers are of the form $0^{c\log\,n}1^{c\log\,n}$, where $c$ is a constant, i.e., markers are concatenations of sufficiently long runs of $0$s and $1$s that are prohibited from occurring within the blocks. The structure of the markers and the property of the code used for the blocks ensure that with high probability, one can split the traces into shorter blocks associated with substrings of length $O(\log^2 n)$, and then run some worst-case trace reconstruction algorithm on the blocks individually. As a result, for every constant deletion probability $q<1$, one can ensure the existence of an efficiently encodable code $\cC\subseteq \{{0,1\}}^{n+r}$ with redundancy $r=O(n/\log n)$ that can be efficiently reconstructed from $\exp(O(\log^{2/3}n))$ traces. Note that reconstruction only requires identifying the markers and reconstructing (in parallel) multiple short-length blocks. This construction can be easily improved while preserving the efficiency of encoding and reconstruction by repeating the process, i.e., making the approach nested. More precisely, we can perform a further partition of all blocks into even shorter subblocks and add a second level of markers: each block of length $\log^2 n$ can be partitioned into blocks of length $(\log\log n)^2$, with markers of length $O(\log\log n)$ added between them. The reconstruction procedure is almost identical to the one already described, except for the fact that a small fraction of blocks will very likely not be reconstructed properly. This issue can be resolved by adding error-correction redundancy to the string to be encoded, resulting in the following claim: for every constant deletion probability $q<1$, there exists an efficiently encodable code $\cC\subseteq \{{0,1\}}^{n+r}$ with redundancy $r=O(n/\log\log n)$ that can be efficiently reconstructed from $\exp(O(\log\log n)^{2/3})$ traces. 

Even this result can be further improved provided that the deletion probability is a sufficiently small constant, in which case modified average-case trace reconstruction algorithms can be used to substantially reduce the number of traces required. This can be achieved with a negligible rate loss. The key idea is that one can efficiently encode $n$-bit messages into strings that are \emph{almost subsequence-unique} via constructions based on \emph{almost $\kappa$-wise independent random variables}~\cite{alon1992simple}. The enabling result for this type of trace reconstruction is the average-case algorithm from~\cite{holenstein2008trace} which is specifically designed to operate on subsequence-unique strings.  

A random vector $X\in\{{0,1\}}^m$ is said to be \emph{$\epsilon$-almost $\kappa$-wise independent} if for all sets of $\kappa$ distinct indices $i_1, i_2,\dots,i_{\kappa}\in\{1,\dots,m\}$ we have
	\begin{equation*}
	|P[X_{i_1}=x_1,\dots,X_{i_{\kappa}}=x_{\kappa}]-2^{-\kappa}|\leq \epsilon,
	\end{equation*}
	for all $(x_1,\dots,x_{\kappa})\in\{0,1\}^{\kappa}$.
In words, we require that every possible $\kappa$-subsequence has probability close to $2^{-\kappa}$ -- the probability distribution induced by any collection of $\kappa$ coordinates of the random vector is close to uniform. Interestingly, such random vectors have explicit constructions based on expander graphs or duals of BCH codes~\cite{alon1992simple,naor1990small}. The trace reconstruction algorithm that exploits the structure of almost subsequence-unique strings relies on an interesting voting strategy that \emph{does not treat every trace equally,} but weighs them according to their reliability. Since analyzing the reconstruction method using probability measures that capture the reliability of traces is difficult, the authors suggest to only use traces that match the last $O(log n)$ already reconstructed bits (as this gives high confidence that the trace is ``synchronized'' with the current estimate). Note that the idea behind this approach is reminiscent of the one used in~\cite{yazdi2017portable}, where the accuracy of traces was estimated based on the accuracy of the address strings that are known to both the information reader and writer. Furthermore, In addition to combining constructions of almost subsequence-unique strings with the corresponding average-case reconstruction algorithm, one also needs to carefully adapt the marker-based approach since the bootstrapping approach used in~\cite{holenstein2008trace} fails for the concatenated runs case. 

With these considerations in mind, one can prove the following results. First, there exists an absolute constant $q^\star>0$ such that for all $q\leq q^\star$ there exists an efficiently encodable code $\cC\subseteq \{{0,1\}}^{n+r}$ with redundancy $r=O(\log n)$ that can be efficiently reconstructed from $\poly(n)$ traces with deletion probability $q$. Second, there exists another absolute constant $q^\star>0$ (to avoid notational clutter, we used the same notation although the constants are different) such that for all $q\leq q^\star$ there exists an efficiently encodable code $\cC\subseteq \{{0,1\}}^{n+r}$ with redundancy $r=O(n/\log n)$ that can be efficiently reconstructed from $\poly(\log n)$ traces with deletion probability $q$. 

Next, we describe how to convert results pertaining to binary codes to codes over larger alphabets. The main claim is that the existence of a binary trace-reconstruction code $\cC$ of length $n$ with rate $R$ that can be efficiently encoded and reconstructed from $t$ traces with error probability $\epsilon$ implies the existence of a $Q$-ary code $\cC'$, where $Q=2^k$, of the same rate $R$. The latter can also be efficiently encoded and reconstructed from $t$ traces with error probability at most $k\epsilon$.

To see this, consider a code whose codestrings are concatenations of binary codestrings of the form
\begin{equation*}
\cC'=\{(c^1,c^2,\dots,c^k)\,:\,c^i\in\cC, i=1,\dots,k\}\subseteq \{{0,1\}}^{k\cdot n}.
\end{equation*}
Clearly, the code $\cC'$ can be viewed as a $Q$-ary code of length $n$ and rate $R$ by encoding the $Q$-ary symbols using the $k$ binary coordinates of the strings $c^i, i \in [k]$. Now, assume that $T'$ is a trace of some 
codestring $c'=(c^1,c^2,\dots,c^k)\in\cC'$. Observe that the trace $T_i$ obtained by replacing each $Q$-ary symbol in $T'$ by the $i$-th bit of its binary expansion has the same distribution as a trace of $c^i$. As a result, applying the transformation $T\mapsto T^i$ to each of the $t$ traces of $c'$ and running the reconstruction algorithm associated with $\cC$ allows us to recover $c^i$ with error probability at most $\epsilon$.

Since this holds for every $i=1,\dots,k$, a union bound over all $i$ shows that we can simultaneously recover $c^1,c^2,\dots,c^k$ from $t$ traces of $c'$ with error probability $\leq k\epsilon$. If the above described constituent binary codes are efficiently encodable, then the $Q$-ary codes are efficiently encodable as well. The codes can also be designed to ensure balanced $GC$-content. To satisfy the balancing constraint, one has to use different markers and a specialized code over the blocks. More precisely, within the blocks, balanced markers of the form $(AC)^\ell \, (TG)^\ell$, are used, with $\ell=25\log n$, where $n$ as before denotes the codelength. Clearly, these markers have the same length as the binary markers used in the two-level marker-based binary code construction.

Consequently, we may summarize the results for $Q$-ary codes as follows. For every constant deletion probability $q<1$, there exists an efficiently encodable code $\cC\subseteq\{A,C,G,T\}^{n+r}$ with redundancy $r=O(n/\log n)$ and balanced $GC$-content that can be efficiently reconstructed from $\exp(O(\log^{2/3}n))$ traces. For every constant deletion probability $q<1$, there exists an efficiently encodable code $\cC\subseteq\{A,C,G,T\}^{n+r}$ with redundancy $r=O(n/\log\log n)$ and balanced $GC$-content that can be efficiently reconstructed from $\exp(O(\log\log n)^{2/3})$ traces.
A summary of the coded trace reconstruction results can be found in Table~\ref{table:results}.

\begin{table*}
	\centering
    \normalsize
	\begin{tabular}{|c|c|c|}
		\hline
		Redundancy             & Number of traces                            & Parameter regime                                           \\ \hline
		$\frac{n}{\log n}$     & $\exp(\log^{2/3}n)$                         & Any constant deletion probability, balanced GC-content \\ \hline
		$\frac{n}{\log\log n}$ & $\exp((\log\log n)^{2/3})$                  &Any constant deletion probability, balanced GC-content \\ \hline
		$\frac{n}{\log n}$     & $\poly(\log n)$& Small enough constant deletion probability             \\ \hline
	\end{tabular}
	\vspace{0.1in}
	\caption{Summary of the properties of trace reconstruction codes. To avoid notational clutter, constants are omitted.}
	\label{table:results}
\vspace{-0.2in}
\end{table*}

We conclude this exposition by referring the interested reader to a \emph{hybrid coded trace reconstruction approach}~\cite{gabrys2017hybrid}, which in addition to traces uses combinatorial families known as $k$-decks: collections of all subsequences of length $k$ of a given string of length $n$.\\

\textbf{3. Set-Codes with Small Discrepancy for DNA Punchcards}\\ 

Code designs for DNA Punchcards are fundamentally different from those used in other molecular storage systems. DNA Punchcards have readily available native sequences that serve as references for alignment of the fragments created via nicking. In all experiments performed on this system (which were of moderate scale), no alignment or readout errors were observed. Consequently, no error-correction was needed to ensure correct reading of the nicking information. However, this type of native DNA-based storage framework suffers from \emph{duplex stability} issues. Stability problems arise when nicks are placed in close proximity to each other, causing disassociation of the DNA fragment straddled by the nicks. Since a nick can be placed either on the $3'-5'$ or $5'-3'$ strand, distributing the nicks in a nearly balanced fashion across the strands is expected to increase duplex stability. Furthermore, if the number of sites actually nicked is small compared to the total number of available nicking sites, the disassociation problem is reduced further. The only conceivable way in which error could occur is to either have defective guides that fail to recognize the correct locations to be nicked or off-target nicking activities. Therefore, requiring further that the combinations of nicked locations of different codestrings differ substantially would resolve this issue as well.

To construct balanced and nonconfusable nick-based codestrings, we will use the notion of \emph{set discrepancy,} introduced in~\cite{beck1981balanced}. Set discrepancy theory has been studied in a number of works~\cite{doerr2003multicolour,lovasz1986discrepancy,muthukrishnan2012optimal}, and has found applications in pseudorandomness and independent permutation generation~\cite{armoni1996discrepancy,saks2000low}, $\epsilon$-approximations and geometry~\cite{matouvsek1993discrepancy}, bin packing, lattice approximations and graph spectral analysis~\cite{doerr2001lattice,rothvoss2013approximating,solymosi2009incidences}.

Informally, the discrepancy of a finite family of subsets over a finite ground set equals the smallest integer $d$ for which the elements in the ground set may be labeled by one of the labels $\pm 1$ so that the absolute value of sums of labels within each subset is at most $d$. In a sense, discrepancy measures how difficult it is to find a labeling of elements that would keep all subsets of the family as close to being balanced as possible.

The formal definition for our storage problem setting is as follows. Let $[n]=\{1,2,\ldots,n\}$. A family of subsets of $[n]$, $\mathcal{F}_n=\{{F_1,\ldots,F_s\}}$, $s\geq 2$, is \emph{$k$-regular} if for all $1 \leq j \leq s$, $|F_j|=k$. Otherwise, the family is {\em irregular}. Let $L: [n] \to \{+1,-1\}$ be a labeling of the elements in $[n]$. The {\em discrepancy} of a set $F_j \in \mathcal{F}_n$ under the labeling $L$ is $D_L(F_j)=\left| \sum_{i \in F_j} L(i) \right|$. The discrepancy of the family $\mathcal{F}_n$ of sets is defined as
$$D(\mathcal{F}_n)=\min_{L} \max_{1 \leq j \leq s} \left| \sum_{i \in F_j} L(i) \right|.$$

For the particular problem of code design for Punchcard systems, we are interested in families of sets $\mathcal{F}_n$ that have \emph{small intersections}, since the sets in the family $\mathcal{F}_n$ are to represent ``codesets'' (i.e., we choose to represent codestrings as sets indicating the locations of $1$s) whose every coordinate is a potential nicking site. By using codesets to represent combinations of nicking sites, it is natural to require that the codesets have small intersections (i.e., the codestrings have largely mismatched locations of $1$s). The codeset formalism also allows for simpler formulations of the coding problems in terms of set discrepancy and set intersection constraints. Although we focus on regular families $\mathcal{F}_n$, there is no inherent reason not to use irregular families as well.  

To this end, we say that the sets in $\mathcal{F}_n$ have {\em $t$-bounded intersections} if for all pairs of distinct integers $i,j \in [s],$ $|F_i \cap F_j|\,< t$. Clearly, for a $k$-regular family $\mathcal{F}_n$, $t < k$ since we do not allow repeated sets. For fixed values of $n$ and $t$, our goal is to find the largest size of a $t$-bounded intersection family $\mathcal{F}_n$ for which there exists a labeling $L$ such that $D_L(F_j)  \in \{-1, 0, +1\}$ for all $1 \leq j \leq s$. We refer to such a set system as an \emph{extremal balanced family}. 

A line of work addressing a similar balanced set-family question in the context of combinatorial designs appeared in~\cite{colbourn1999bicoloring}. More precisely, the problem studied is \emph{bicoloring} of Steiner triple systems (STSs). Roughly speaking, Steiner triple systems are set systems in which the subsets of interest satisfy intersection constraints that ensure that each pair of distinct elements of the ground set appears in exactly one subset (block) of the system. The key finding is that STSs are not perfectly bicolorable, i.e., that there will always exist a monochromatic triple. 

To design extremal balanced families, one can start with known families of sets with small intersections, such as the Bose-Bush and Babai-Frankl families~\cite{babai1992linear,bose1952orthogonal}. In this case, one can achieve the smallest possible discrepancy ($d=0$ for even-sized sets and $d=1$ for odd-sized sets) in a natural manner, 
by using only the defining properties of the sets. 

Let $q$ be a prime power such that $1 \leq t \leq k \leq q$, and let $n = kq$. Furthermore, let $\xi$ be a primitive element of the finite field $\fq$. Set $\A = \{0,1,\xi,\ldots,\xi^{k-2}\}$. Clearly, $|\A| = k$. For each polynomial $f \in \fq[x]$, define a set of ordered pairs of elements from the underlying finite field:
\[
A_f \define \{(a,f(a)) \colon a \in \A\}.
\]
Clearly, $|A_f| = k$. Furthermore, let
\begin{equation} \label{eq:babai}
\C(k,q) \define \{A_f \colon f \in \fq[x], \deg(f) \leq t-1\}.
\end{equation}
Then $\C(k,q)$ is a family of $q^t$ $k$-subsets of the set $\X \define \A \times \fq$ such that every two sets intersect in at most $t-1$ elements. This follows because two distinct polynomials of degree $\leq t-1$ cannot intersect in more than $t-1$ points. The Ray-Chauduri and Wilson Theorem~\cite{babai1992linear} asserts that the size of any family $\mathcal{F}_n$ of $k$-regular sets with $k \geq t$ whose pairwise intersection cardinalities lie in some set of cardinality $t$ satisfies $|\mathcal{F}_n| \leq \binom{n}{t}$. As an example, the set of all $t$-subsets of $[n]$ forms a $(t-1)$-intersection bounded $t$-regular family of subsets. Under certain mild parameter constraints, the aforementioned result can be strengthened when the set of allowed cardinalities equals $\{{0,1,\ldots,t-1\}}.$ The size of the family is roughly $\frac{n^t}{k^t}$.

Given the simple definition in~\eqref{eq:babai}, one can easily devise a labelling $L$ 
of the pairs of points $(a,f(a))$ such that every set in the family $\C(k,q,s)$ has discrepancy $=0$, for even $k$, and discrepancy $=\pm 1,$ for odd $k$. For completeness, we next present the simple proof of this claim from our work~\cite{gabrys2020set}. 

The first step consists of disposing of the representation of a point in terms of a pair of symbols from the underlying finite field. To this end, let a map $M$ operating on $\fq$ be such that $M(0) = 0$ and $M(a) = m+1$ if $a = \xi^m \neq 0$. It is easy to check that $M(a) \in [0,k-1],$ $\forall \, a \in \A$ and
$M(b) \in [0,q-1],$ $\forall \, b \in \fq$. The pair $(a,b) \in \X = \A \times \fq$ is mapped to $\sigma(a,b) = qM(a) + M(b) \in [0,n-1]$, and $M$ is a bijection.

Assume that $k$ is even. We claim that for every set $A_f$, half of the elements are mapped to $[0,n/2-1]$ while the other half of the elements are mapped to $[n/2,n-1]$. To see why this is true, note that for $a \in \{0,1,\xi,\ldots,\xi^{k/2-2}\} \subset \A$ and $b = f(a) \in \fq$, the pair $(a,b)$ is mapped to 
\[
\sigma(a,b) = qr(a) + r(b) \leq q(k/2-1) + (q-1) = n/2-1. 
\] 
Similarly, for $a \in \{\xi^{k/2-1},\ldots,\xi^{k-2}\} \subset \A$ and $b = f(a)$, we have
\[
\sigma(a,b) = qr(a) + r(b) \geq qk/2 + 0 = n/2. 
\]
Based on this result, one can perform a labeling as follows. Assign $-1$ to $(a,b)$ if $\sigma(a,b) < n/2$, and assign $+1$ to $(a,b)$ if $\sigma(a,b) \geq n/2$. Then, every set in the family $\C(k,q,s)$ has half of the elements mapped to $-1$ and half mapped to $+1$. Equivalently, the discrepancy of every set is equal to $0$. The case when $k$ is odd can be handled similarly.

Three remarks are in place. First, the balancing property directly follows from partitioning the set $\A$. Second, the construction of the sets is reminiscent of the well-known and ubiquitous Reed-Solomon construction. Third, the already mentioned connection of the coding problem to combinatorial design theory suggests other constructions, and throughout the remainder of the section we focus on one such approach based on \emph{transversal designs.} 

A {\em transversal design}~\cite{colbourn2010crc} $TD(t,k,v)$ consists of a set $V$ of $kv$ elements, called points,  a partition of $V$ into sets 
$\{ G_i : i \in [k] \},$ called groups. The sets $G_i, i \in [k],$ contain $v$ points. In addition, we have a set $\B$ of $k$-subsets called blocks. A block and a group obey intersection constraints that can be summarized as follows: every $t$-subset of $V$ is either contained in exactly one block or in exactly one group, but not both. Because no $t$-subset of elements can appear in two or more blocks, any two distinct blocks of a $TD(t,k,v)$ intersect in at most $t-1$ elements. Therefore, whenever a $TD(t,k,v)$ exists, one can use it to construct a family of sets with small intersections that are simultaneously balanced by mimicking the proof described above. To summarize, we assign $+1$ labels to the points in half of the groups and $-1$ labels to the points in the other half of the groups when $k$ is even (and follow a similar approach for odd $k$). It is straightforward to see that the Bose-Bush/Babai-Frankl construction actually represents a transversal design, which was first pointed out in~\cite{stinson1981general}. Furthermore, it is not difficult to add $k$-blocks to the design and still retain the balancing and intersection constraints.

For simplicity, assume that $k$ is even and that $t \geq 3$. There has to be one group of the TD that is completely contained within the set of positively labeled elements (which we henceforth denote by $P_{+}$), and one group that is completely contained within the set of negatively labeled elements (which we henceforth denote by $P_{-}$). There are $\left(\frac{k}{2}\right)^2$ such pairs of groups. By construction, any $k$-subset with $\frac{k}{2}$ points from the first group and $\frac{k}{2}$ points from the second group intersects each block of the TD in at most two points. Furthermore, each pair of such blocks intersects in at most $\lceil \frac{k}{2} \rceil$ points. Hence, if $t> \max\{{\frac{k}{2},2\}}$, the blocks used to augment the TD are both balanced and satisfy the required intersection constraint. This construction easily extends to larger selections of groups. In this case, as long as $t > \max\{{\frac{k}{s}, s\}}$, where $s$ is the size of the collection of groups, the new blocks satisfy the required constraints provided that any two collections of $s$ groups have fewer than $\frac{t}{k/s}$ groups in common. An important conclusion that arises from this argument is that TD and related combinatorial designs may not directly lend themselves to extremal balanced families of sets with small intersections. Instead, one may need to use combinations of designs and augment them, and several such constructions based on mutually orthogonal Latin squares and derivatives of orthogonal arrays and packings have been reported in~\cite{gabrys2020set}.\\

\textbf{4. Coding for Unique Reconstruction}\\ 

The three different coding-theoretic problems pertaining to DNA error-correction and constrained coding for shotgun and nanopore sequencing, as well as DNA Punchcard systems, do not deal with another fundamental class of problems we termed \emph{unique string reconstruction}. With the constraints imposed by individual sequencing devices on the type of outputs produced, one of the most important questions is to ensure that even in the ideal case of no sequencing errors, a DNA string can be uniquely reconstructed from the available output data of the sequencer. 

To illustrate this requirement, consider the following example of two distinct binary strings, $\bfx=10010$ and $\bfy=00100$. 
Let $S_L(\bfz)$ denote the set of all substrings of the string $\bfz$ of length $L$. Then, $S_3(\bfx)=S_3(\bfy)=\{{100,001,010\}},$ and based on the substring information alone, one cannot distinguish the strings $\bfx$ and $\bfy$. Increasing $L$ from $3$ to $4$ leads to $S_4(\bfx)=\{{1001,0010\}}\neq S_4(\bfy)=\{{0010,0100\}}.$ Therefore, based on the two substrings of length $4$, one can discriminate the two possible (input) strings. A general result regarding the uniqueness of string reconstruction based on substrings of length $L$ was derived in~\cite{ukkonen1992approximate}, where it was shown that a string is uniquely $L$-substring reconstructable if all its $L-1$-substrings occur at most once (i.e., if there are no repeats). Another important result in the area~\cite{skiena1995reconstructing,margaritis1995reconstructing} established that unique $L$-substring reconstruction is impossible for strings with period $\rho \leq L$ (a string $\bfx$ is said to have period $\rho$ if $x_i=x_{i+\rho}$, for all $1 \leq i \leq n-\rho$). Otherwise, $L \geq \lfloor n /2 \rfloor+1$ suffices for unique reconstruction. For example, $S_4(0111011)=S_4(1110111)=\{{0111,1110,1011,1101\}}$, since $\rho=4$ and $L=4$.

Native (mammalian) DNA usually contains a large number of repeats~\cite{jelinek1980ubiquitous} and as a result, modern sequencing technologies are being redesigned to produce long reads~\cite{huddleston2014reconstructing} that can use the context of the repeats to ensure unique reconstruction. Adapting the sequence content for ease of reconstruction is, in this case, obviously impossible. But once again, that is not the case for DNA-based data storage applications, since one can encode the strings to avoid repeats, as first suggested in~\cite{gabrys2019unique}. The problem addressed in~\cite{gabrys2019unique} can be summarized as follows. Let $\mathcal{C}_L$ be a set of binary codestrings $\bfx$ of length $n$, each of which can be uniquely reconstructed based on $S_L(\bfx)$. What is the largest size of $\mathcal{C}_L$ for a given $L$ and can the code(s) be efficiently encoded and decoded? The question was addressed affirmatively, establishing that codes $\mathcal{C}_L$ of asymptotic rate equal to $1$ (i.e., with only a constant number of redundant bits) exist whenever the substrings are long enough, i.e., $L>2 \log(n)$. These codes can be encoded using a specialized \emph{repeat-removal} procedure, which replaces repeats with pointers to the locations of their first occurrence, reminiscent of but significantly more involved than a related procedure for runlength coding~\cite{van2010construction}.  

Other relevant coding methods for unique string reconstruction include~\cite{acharya2015string,gabrys2020mass,pattabiraman2023coding,gabrys2022reconstruction}. There,
strings are reconstructed based on masses (i.e., weights) of their substrings, or prefixes and suffixes only, without knowing the actual substrings themselves. This subsequence-weight reconstruction problem is motivated by \emph{mass spectrometry sequencing}~\cite{chen2020bioinformatics} and its application to data storage in synthetic polymers~\cite{laure2016coding}. The interested reader is referred to the original manuscripts for an in-depth coverage of the topics, with solutions including mixtures of ideas from the area of the turnpike reconstruction problem~\cite{dakic2000turnpike}, code constructions based on Catalan strings and modifications thereof~\cite{stanley1999exercises} and binary $B_h$ sequences~\cite{gabrys2022reconstruction} and constant-weight codes~\cite{sima2023constant}. 

%\section{Open Problems} \label{sec:conclusions}
\clearpage
\bibliographystyle{IEEEtran}
\bibliography{IEEEabrv,hanmao,ref-coding,ref-olgica,ref-hossein}

\end{document}